\documentclass[aps,amsmath,twocolumn,a4paper,prb]{revtex4-1}

\usepackage{graphicx}
\usepackage{amssymb}
\usepackage{amsmath}

\newcommand {\be} {\begin{equation}}
\newcommand {\ee} {\end{equation}}
\newcommand {\bea} {\begin{eqnarray}}
\newcommand {\eea} {\end{eqnarray}}
\newcommand {\bes} {\begin{displaymath}}
\newcommand {\ees} {\end{displaymath}}
\newcommand {\beas} {\begin{eqnarray*}}
\newcommand {\eeas} {\end{eqnarray*}}

\newcommand*{\lss}{\ell_{\rm ss}}
\newcommand*{\Nss}{N_{\rm ss}}
\newcommand*{\lds}{\ell_{\rm ds}}
\newcommand*{\Nds}{N_{\rm ds}}

\newcommand*{\lpa}{\left(}
\newcommand*{\rpa}{\right)}
\DeclareMathOperator\erfc{erfc}

\begin{document}

\title{How nanochannel confinement affects the DNA melting transition within
  the Poland-Scheraga model}

\author{Michaela Reiter-Schad} 
\affiliation{Department of Astronomy and Theoretical Physics, 
Lund University, S\"olvegatan 14A, Lund, SE-223 62 Lund, Sweden}

\author{Erik Werner} 
\affiliation{Department of Physics University of
  Gothenburg Origov\"agen 6B SE-412 96 G\"oteborg, Sweden}

\author{Jonas O. Tegenfeldt}
 \affiliation{Division of Solid State Physics,
  Department of Physics, Lund University, 
Box 118, SE-221 00 Lund, Sweden}

\author{Bernhard Mehlig}
\affiliation{Department of Physics University of
  Gothenburg Origov\"agen 6B SE-412 96 G\"oteborg Sweden.}

 \author{Tobias Ambj\"ornsson}
\affiliation{Department of Astronomy and Theoretical Physics, Lund University,
  S\"olvegatan 14A, Lund, SE-223 62 Lund, Sweden.}

\begin{abstract}
  When double-stranded DNA molecules are heated, or exposed to denaturing
  agents, the two strands get separated. The statistical physics of this
  process has a long history, and is commonly described in term of the
  Poland-Scheraga (PS) model. Crucial to this model is the configurational
  entropy for a melted region (compared to the entropy of an intact region of
  the same size), quantified by the loop factor. 
  In this
  study we investigate how confinement affects the DNA melting transition, by
  using the loop factor for an ideal Gaussian chain. By subsequent
  numerical solutions of the PS model, we demonstrate that the melting
  temperature depends on the persistence lengths of single-stranded and
  double-stranded DNA. For realistic values of the persistence lengths the
  melting temperature is predicted to decrease with decreasing channel
  diameter. We also demonstrate that
  confinement broadens the melting transition.  These general findings hold
  for the three scenarios investigated: namely 1. homo-DNA, i.e. identical
  basepairs along the DNA molecule; 2. random sequence DNA, and 3. ``real''
  DNA, here T4 phage DNA. We show that cases 2 and 3 in general give rise to 
  broader transitions than case 1.  Case 3 exhibits a similar phase transition
  as case 2 provided the random sequence DNA has the same ratio of AT to GC
  basepairs. A simple analytical estimate for the shift in melting temperature
  is provided as a function of nanochannel diameter. For homo-DNA, we also
  present an analytical prediction of the melting probability as a function of
  temperature.

\end{abstract}

\date{\today}


\maketitle

\section{Introduction}

The three dimensional structure of double-stranded Deoxyribonucleic acid
(dsDNA) DNA is the famed Watson-Crick double-helix, within a broad range of
salt and temperature conditions.\cite{watsoncrick} This double-helix encodes
the genetic information of an entire organism in terms of four types of
bases. Base complementarity guarantees that the same information is contained in
the two strands of a dsDNA.\cite{alberts}

\begin{figure}
\begin{center}
\includegraphics[width=8cm]{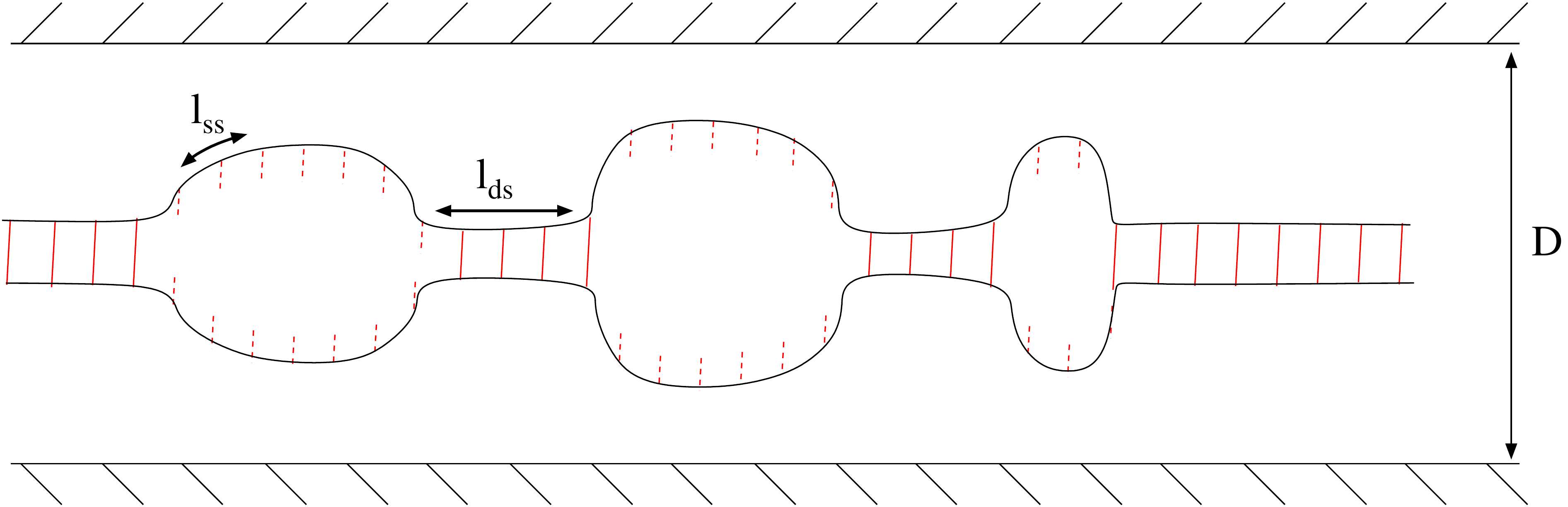}
\end{center}
\caption{Cartoon of DNA melting in a nanochannel. A given microstate of DNA,
  at elevated temperatures, is an alternating set of intact double-stranded
  regions and melted single-strand regions (DNA bubbles). Within the
  Poland-Scheraga model, the statistical physics of DNA melting is described
  in terms of two hydrogen-bond energies, ten stacking (nearest neighbor)
  parameters and the loop factor $g(m)$ which quantifies the conformational
  entropy for a melted region corresponding to $m$ basepairs (relative to the
  entropy of an unmelted region of the same size). The functional form of
  $g(m)$ is different for unconfined and nanoconfined DNA. }
\label{fig:schematic}
\end{figure}
By temperature increase the double-stranded double-helical DNA progressively
denatures -- DNA melting, see Figure \ref{fig:schematic}. Partially melted DNA
is an alternating set of intact, stiff, double-stranded regions (persistence
length about 50 nm) and of single-stranded, floppy (persistence length
about 1 to a few nm), ``melted'' regions (DNA bubbles).\cite{poland} Each of
these regions is subject to thermal fluctuations. Interestingly, the average
free energy associated with breaking an AT-basepair is smaller than the
corresponding energy for a GC-basepair.\cite{gueron} In melting studies,
therefore, DNA tends to melt first in AT-rich regions ($45.79$ $^\circ$C at 
salt 
concentration 0.01 M for pure AT) and only at higher temperature 
($95.88$ $^\circ$C at salt concentration 0.01 M for pure GC regions) 
in GC-rich regions.

On the theoretical side, The Poland-Scheraga (PS) model of DNA melting has
been proven to well reproduce (macroscopic) melting
data.\cite{poland,yeramian,santalucia,FKreview,wartell,blake,blossey} Note,
however, that there is an alternative model, the Peyrard-Bishop 
model,\cite{peyrard,Kalosakas_Rasmussen} see Ref. \onlinecite{FKreview} for a
comparative study of the two models. Herein we use the PS model, which is an
Ising model with a long-range term due to the entropy associated with the
melted single-stranded regions, and has the following parameters: two hydrogen
bond (Watson-Crick) energies (AT or GC bonds), ten independent stacking
(nearest neighbor) parameters, the loop factor (or, loop function\cite{blake}), 
and the ring-factor
(bubble initiation) parameter (related to the cooperativity
parameter).\cite{poland} Rather recently the ten stacking parameters were
independently measured for the first time for different temperatures and salt
concentrations.\cite{FK}

Theoretical interest in the melting transition originates, in part, from the
fact that the PS model exhibits a phase-transition which can be analyzed
exactly.\cite{fisher_84,Wiegel} This analysis treats {\it unconfined} DNA
molecules, with {\it identical} hydrogen bond and stacking energies
(homo-DNA). For unconfined DNA the loop factor $g(m)$ (see above)
scales with DNA bubble size $m$ as a power-law: $g(m) \simeq m^{-c}$ with a
loop exponent $c$. The exponent $c$ determines the order of the phase
transition; if $c>2$ the melting transition is first order.\cite{kafri}

More recent studies\cite{hwa,garel_04,giacomin2009,alexander2009,derrida2014}
address the challenge of understanding how the DNA sequence affects the
melting transition, typically by analyzing the melting of random sequence
DNA.\cite{derrida_83,azbelI,azbelII,azbelIII,lifshitz} Unlike for the case of
homo-DNA melting, no exact results exist for these kinds of systems.

Herein, we go beyond previous studies by considering how confinement affects
the DNA melting transition. In particular, we seek to quantify how confinement
affects the DNA melting temperature and the width of the melting transition. In
contrast to Ref. \onlinecite{li_2014}, where molecular dynamics simulations of
very short DNA (15 basepairs) were performed, we are here interested in the
thermodynamic limit (we study typical DNA sizes of $2\cdot 10^5$ basepairs).  In
Ref.~\onlinecite{werner_schad2015} we calculate how confinement influences
$g(m)$ and show how this influences melting for a highly simplified model
system.  In the present study, we go beyond Ref.~\onlinecite{werner_schad2015}
by calculating what consequences the effect of confinement on $g(m)$ has for
the melting of a more realistic model of DNA.  Using as input the functional
form for $g(m)$,\cite{werner_schad2015} we provide results of extensive
simulations within the Poland-Scheraga model for 1. homo-DNA, 2. random
sequence DNA and 3. ``real'' DNA (T4 phage). We also provide a simple formula
for estimating how the melting temperature changes as a function of the
channel diameter and present an analytical prediction of the melting
probability of homo-DNA as a function of temperature.

The present study is inspired by recent experiments which demonstrate the
potential to study local properties of the DNA melting transition in
nanochannels.\cite{reisner2010,reisner2012} Other recent experiments now allow
studying also the DNA melting dynamics.\cite{altan,our_prl}

The biological relevance of DNA melting in confined environments is, for 
instance, due to the fact that
local denaturation of DNA is necessary for protein binding to a DNA
single-strand,\cite{mark,mark1} and is implicated in transcription
initiation.\cite{kalo,yeramian}

\section{Review of the Poland-Scheraga model}\label{sec:recapitulation}

In this section we introduce the Poland-Scheraga (PS) model for unconfined
DNA. In the next section we discuss how the PS model must be modified in order 
to study
DNA melting under confinement.

\subsection{General considerations}\label{sec:general}
 
Consider double-stranded DNA with $N$ internal basepairs, that are clamped at
both ends for simplicity (Fig.~\ref{fig:schematic}). The free energy required
for breaking a hydrogen bond at basepair $i$ is denoted by $E_{\rm
  hb}(i)$. There are two values for this parameter (AT-bonds or GC-bonds). For
disrupting the stacking (nearest neighbor) interactions between basepairs
$i-1$ and $i$ there is a free energy cost $E_{\rm st}(i-1,i)$. There are ten
stacking parameters.\cite{FK} In addition we need the ring factor $\xi\approx
10^{-3}$ (see Ref.~\onlinecite{FK}) which is a Boltzmann factor associated
with the free energy cost of creating two ``boundaries'' between intact and
melted DNA.  Note that $E_{\rm hb}(i)$ and $E_{\rm st}(i-1,i)$ have energetic
as well as entropic contributions: $E_{\rm hb}(i)=U_{\rm hb}(i) - T S_{\rm
  hb}$ and $E_{\rm st}(i-1,i) = U_{\rm st}(i-1,i) - T S_{\rm st}$, where $U$
denote energetic contributions, $T$ is the temperature and $S_{\rm hb}$ and
$S_{\rm st}$ are entropies, commonly assumed to be basepair independent.
Finally, the Poland-Scheraga model requires as input the ratio of the number
of configurations for a melted region (DNA bubble) and the number of
configurations of an intact region, quantified by the loop factor $g(m)$,
see sections \ref{sec:loop_factor_unconfined} and
\ref{sec:loop_factor_confined}.

\subsection{The melting temperature and the Marmur-Doty formula} 
\label{sec:Marmur-Doty}

Throughout the main text we repeatedly refer to the melting temperature,
$T_M$, or deviations therefrom. $T_M$ is defined as the temperature where the 
melting probability = 1/2.

We can estimate the melting temperature, $T_M$ by setting the 
total free energy per 
basepair, $E_{\rm hb} + E_{\rm st}$, equal to zero. This leads to

\be\label{eq:TM_melt}
T_{\rm M} = (U_{\rm hb}+U_{\rm st})/S_{\rm ref}
\ee
with $S_{\rm ref}=S_{\rm hb}+S_{\rm st}$, and we find that the free energy can 
also be expressed as

\be\label{eq:free_energy}
E = S_{\rm ref} (T_{\rm M} - T).
\ee

The total free energy of a given DNA sequence can be estimated as the sum of 
the free energies of the AT and the GC portion:
\bea\label{eq:F_T_MD}
E&=& f_{\rm AT} \Delta E_{\rm AT} + f_{\rm GC} \Delta E_{\rm GC} \nonumber\\ 
&=& S_{\rm ref} (f_{\rm AT} T_{\rm AT} + f_{\rm GC} T_{\rm GC} - T)
\eea
with the free energy of random AT and random GC sequences $\Delta E_{\rm AT}$ 
and $\Delta E_{\rm GC}$.

With Eq. (\ref{eq:free_energy}) we find the Marmur-Doty formula, which gives a 
simple estimate for the melting temperature, 
$T_M$ of
unconfined DNA: 
\be\label{eq:T_MD}
T_{\rm MD} = f_{\rm AT} T_{\rm AT} + f_{\rm GC} T_{\rm GC}
\ee
where $f_{\rm AT/GC}$ is the fraction of AT/GC bonds in the DNA
sequence. $T_{\rm AT/GC}$ is the melting temperature for pure random AT/GC
sequences. We note that this formula is rather crude, in reality $T_{\rm M}$
depends weakly also on $\xi$ and $g(m)$.

We will later use the stability parameters introduced in
Sec. \ref{sec:general} from Ref. \onlinecite{FK} (Table 1). Herein, the ten
stacking parameters were experimentally determined and the hydrogen bond
energies were found from the empirical Marmur-Doty melting temperature together
with
Eq. (\ref{eq:free_energy}) and \be E_{\rm hb}^{\rm AT} = \Delta E_{\rm AT}
-\frac{1}{4} \sum_{AT,TA,AA,TT} E^{\rm st} \ee and the corresponding equation
for GC. Note that we use a different notation than as Ref. \onlinecite{FK}, $E = 
-
\Delta G$ and $S_{\rm ref} = -\Delta S$.

For homo-DNA, the melting temperature is estimated as: 
\bea\label{eq:TM_A}
T_{\rm M} ^{\rm A} &=& \frac{U_{\rm st}^{\rm AA}+U_{\rm hb}^{\rm 
AT}}{S_{\rm hb}+S_{\rm 
st}} \\ &=& T_{\rm AT} + \frac{1}{4 S_{\rm ref}} (U_{\rm st}^{\rm 
AT} + U_{\rm st}^{\rm TA} - 2 U_{\rm st}^{\rm AA}).
\eea
At salt concentration 0.01 M, we have $T_{\rm M}^{\rm A} = 52.73 ^\circ$C.

Even though investigation
of unconfined DNA is not the main purpose of this study, we elaborate a bit on
deviations from the Marmur-Doty formula in appendix \ref{sec:unconfined_DNA}, 
where we also
investigate the properties of the DNA melting temperature on the value of
$c$ for unconfined DNA.

\section{The loop factor}

In this section we give details about the loop factor for confined and 
unconfined DNA molecules.

\subsection{Loop factor for unconfined DNA}\label{sec:loop_factor_unconfined}

Let us now consider $g(m)$ for unconfined DNA (see also
Ref. \onlinecite{werner_schad2015}).  For simplicity, assume that the
melted DNA region is described by a self-avoiding random walk of $2m$ steps in
3 dimensions on a lattice. If we denote by $z$ the coordination number (number
of nearest neighbors in the absence of self-exclusion), and limit ourselves
to large number of steps, the number of
configurations for a closed loop (ring polymer) is 
$\Omega_{\rm ring}=z^{2m}m^{-c}$, where $c=1.76$.
For a linear polymer (intact DNA),
similarly $\Omega_{\rm linear}=\tilde{z}^{m}$, with a different coordination
number $\tilde{z}$, since the persistence length is in general different for
double-stranded DNA compared to single-stranded DNA. The prefactor
$(z/\tilde{z})^{2m}$ is included in formalism by a redefinition of the the
Watson-Crick energy $E_{\rm hb} \rightarrow E_{\rm hb} + 2 R T
\log(z/\tilde{z}^{1/2})$ and the loop factor then becomes
\be\label{eq:gm_unconfined}
g_u(m)\sim m^{-c}
\ee
for unconfined DNA. If one includes self-avoidance between intact DNA regions
and melted regions one finds $c=2.12$\cite{kafri}. For an unconfined ideal
(phantom) chain one obtains $c=3/2$. A few words of caution are here
needed. Strictly speaking, the quantity $g(m)$ is proportional to the number
of configurations of a random walk of $2m$ steps with the same start and end
point, where the random walk is constrained to have {\em no bound
  complementary basepairs} along its path.\cite{werner_schad2015} 
  We here follow the approximation
common to the literature,\cite{kafri} namely, assuming that $g(m)$ can be
estimated by removing the constraint of no bound states. In
Ref. \onlinecite{werner_schad2015} we discuss the effects caused by including
the constraint mentioned above. 

The power-law form for $g_u(m)$ gives rise to
an effectively long-range interaction for molten DNA 
regions.\cite{blake,fixman,richard} Note that the persistence length of
single-stranded DNA is 1-5 nm. So, the loop-correction $g(m)\sim m^{-c}$
should not be included for small bubbles, although it is common practice to
include it for all $m$-values. A weight of $\xi g(m)$ is assigned to each
bubble. Therefore the value of the ring factor $\xi$ from Ref. \onlinecite{FK},
which was obtained from very small DNA molecules neglecting the
loop-correction, has to be complemented by a loop factor $g_u(m) = m^{-c}$.
We note that a useful form for $g(m)$ approximately correct for all $m$ is 
given in, for instance, Ref. \onlinecite{wartell}.

\subsection{Loop factor for DNA confined in a nanochannel}
\label{sec:loop_factor_confined}

The aim of this study is to understand how channel confinement influences the
melting transition for real DNA. To that end we need to find a functional form
for $g(m)$ for a polymer confined to a nanochannel. In
  general we have:
\begin{equation}\label{eq:gm_confined}
g_c(m) = \frac{G_{\rm ss}(\Nss)}{G_{\rm ds}(\Nds)}
\end{equation}
The quantity $G_{\rm ds}(\Nds)$ is Green's function for a double-stranded DNA
consisting of $\Nds$ Kuhn lengths in a square channel with side length $D$ and
random initial positions for the polymer start position. Similarly, $G_{\rm
  ss}(\Nss)$ is proportional to Green's function for a single-stranded DNA
region of $\Nss$ Kuhn lengths with identical start and end positions [below we
determine the constant of proportionality using known asymptotic results for
$g(m)$]. The number of base pairs is denoted by $m$ and related to the Kuhn
lengths according to $\Nss=2m/(x\lss)$ and $\Nds=m/(x\lds)$ with a conversion
factor $x=3 {\rm bp/nm}$ (the center-to-center distance between adjacent 
basepairs
is $0.34$ nm, see Ref. \onlinecite{alberts}). We use $\lss=6{\rm nm}$ for 
single-stranded DNA and
$\lds=100{\rm nm}$ for double-stranded DNA throughout this study. Determining
$G_{\rm ds}(\Nds)$ and $G_{\rm ss}(\Nss)$ poses the formidable, unsolved, task
of deriving an expression for the number of conformations for a worm-like
chain model\cite{khokhlov} with two constraints: (i) no polymer conformations
can ``pass'' the channel walls, and (ii) the polymer conformations cannot
self-intersect. 
In order to provide expressions
for $G_{\rm ss}(\Nss)$ we neglect
self-avoidance, i.e., the constraint (ii) above.
Notice that for typical experimental channel sizes, $D\sim 50 -100$
nm,\cite{reisner_pedersen_2012} thus $\lss/D\ll 1$. Single-stranded regions 
are therefore well
described by the statistics of an ideal Gaussian chain, and 
$G_{\rm ss}(\Nss)$ is given by the solution to a diffusion 
equation.\cite{khokhlov,werner2013}
The result is given in the Supplementary
Information of Ref. \onlinecite{werner_schad2015}, where we find that for an
ideal Gaussian chain we have
\begin{align}\label{eq:Gss}
 G_{\rm ss}(\Nss)& = A  \Nss^{-1/2} \left[\sum_{k=1}^ 
\infty\exp\lpa- \Lambda^{\rm ss}_k \Nss \rpa\right]^2,\\ 
\Lambda^{\rm ss}_k&  = \frac{\lss^2\pi^2k^2}{6D^2}.
\end{align} 
where $\Lambda^{\rm ss}_k$ are the eigenvalues of the diffusion operator. The 
constant of proportionality, $A$, is obtained below. 

For deriving an expression for $G_{\rm ds}(\Nds)$ we notice that, in
contrast to $\lss/D$, the ratio of double-strand Kuhn length to channel
diameter, $\lds/D$, is not necessarily a small number for realistic channels
sizes. To address this issue we write
\begin{equation}\label{eq:GDs_corrected}
G_{\rm ds}(\Nds)= \gamma (\Nds) G^{\rm diff}_{\rm ds}(\Nds)
\end{equation}
where
\begin{align}\label{eq:GDs}
G^{\rm diff}_{\rm ds}(\Nds) & =   \left[\frac{8}{\pi^2}\sum_{k=0}^ 
\infty\frac{1}{(2k+1)^2} \exp\lpa-\Lambda^{\rm ds}_{2k+1} \Nds \rpa\right]^2,  
\\ 
 \Lambda^ {\rm ds}_k & = \frac{\lds^2\pi^2k^2}{6D^2}. 
\end{align}
The quantity $G^{\rm diff}_{\rm ds}(\Nds)$ is the Green's function for an
ideal Gaussian chain in weak confinement ($D\gg\lds$).\cite{werner_schad2015}
The factor $\gamma (\Nds)$ quantifies deviations between the true statistics
for a double-stranded region and the ideal Gaussian case. In order to estimate
$\gamma(\Nds)$ we note that in Ref. \onlinecite{tree_2013} an interpolation
formula for the free energy of confinement was given which encompasses both
the Odijk regime and the diffusive regime of an ideal worm-like chain.  We
here choose $\gamma(\Nds)$ so that the free energy of confinement, $R \log
[G_{\rm ds}(\Nds )]$, equals the one given in Eq. (13) of
Ref. \onlinecite{tree_2013}, in the limit $\Nds\to \infty$. We then have
\begin{align} 
\gamma(\Nds) &= \exp \lpa 2\Lambda^ {\rm ds}_1 \lpa1-\kappa(\lds/D)\rpa \Nds
\rpa ,  \\
\kappa(z) &= \lpa 1 + 1.672 z + 1.287z ^2 \rpa^{-2/3}. \label{eq:kappaEquation}
\end{align}
In the derivation of the above expressions, we have assumed that the entropy of
an intact or melted section in the interior of a polymer is the same as that
of an isolated polymer of the same size. We briefly discuss the effect of this
assumption in the Supplemental Material of Ref. \onlinecite{werner_schad2015}.

Our final task is now to determine  the constant $A$ in Eq. (\ref{eq:Gss}). To
that end we require that $g_c(m)$ agrees with
the unconfined case for a ideal chain, $g_u(m)=m^{-3/2}$, as $D \rightarrow
\infty$. In order to analyze the large $D$ limit of $g_c(m)$,
Eqs. (\ref{eq:Gss}) and (\ref{eq:GDs}) are not suitable. We therefore rather
use the equivalent resummed expressions 
\begin{widetext}
\begin{equation}\label{eq:Gss_resummed}
G_{\rm ss}(\Nss) = A  \Nss^{-1/2} \left[\sqrt{ 
\frac{6D^2}{\lss^2 \pi \Nss}}\left(\sum_{n=1}^ 
\infty\exp\left(-\frac{6 D^2 
n^2}{\lss^2 \Nss}\right)+\frac{1}{2}\right)
-\frac{1}{2}\right]^2
\end{equation}

\begin{equation}\label{eq:GDs_resummed}
G^{\rm diff}_{\rm ds}(\Nds) =\left[1-\frac{2}{\sqrt{B}}  
\lpa \frac{1}{\sqrt{\pi}} +2 \sum_{n=1}^\infty (-1)^n 
\left\{ \sqrt{\pi} \exp\lpa-Bn^2\rpa - 
\sqrt{B}n \erfc\lpa 
\sqrt{B}n \rpa \right\} \rpa \right]^2 \quad 
\textrm{with} \quad B=\frac{3D^2}{2\lds^2\Nds}
\end{equation}
\end{widetext}
where Eq. (\ref{eq:Gss_resummed}) was obtained using the Poisson resummation
formula, and Eq. (\ref{eq:GDs_resummed}) is derived in
Ref. \onlinecite{carslaw} (chap. 3). With $\lim_{D \rightarrow \infty}
G_{\rm ds}(\Nds) = 1$ we find
\begin{equation}
A=\frac{2\pi}{3} \lpa\frac{2}{x\lss}\rpa^{3/2}\lpa\frac{\lss}{D}\rpa^2
\end{equation}
With this result, the known value of the ring factor for free DNA $\xi=
10^{-3}$ can be used.

\begin{figure}
    \hspace{-0.7cm}\includegraphics[width=9cm]{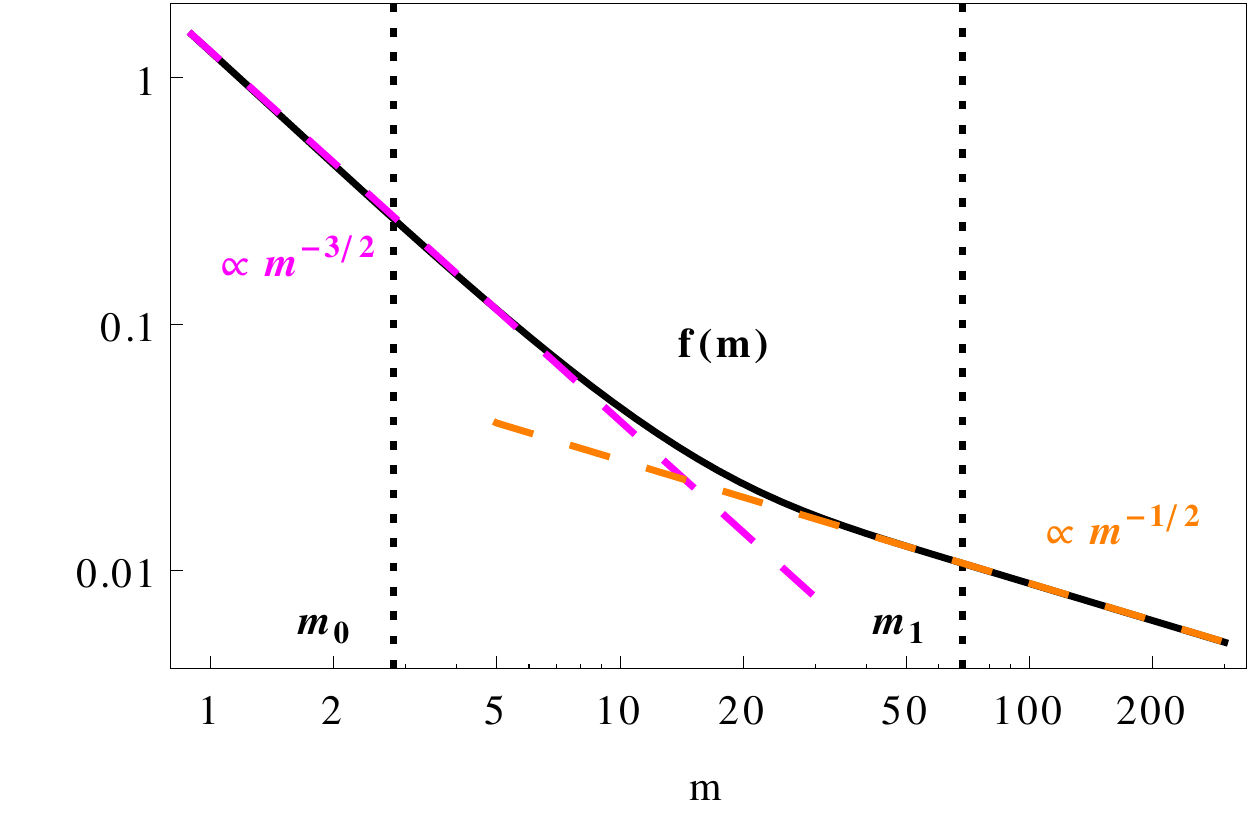}
    \caption{Shifted loop factor $f(m)$ for a polymer confined to
      a square channel of diameter $D$. The black solid curve shows the exact 
      result for an ideal chain contained in Eqs.
      (\ref{eq:gm_confined})--(\ref{eq:GDs_resummed}). In general we
      identified three regimes separated by transition points $m_0$ and $m_1$
      (see main text). For small and large $m$ we have that a power-law
      behavior, $f(m)\propto m^{-3/2}$ and $f(m)\propto m^{-1/2}$,
      respectively.  For illustrative purposes we used a small
        value for the ratio of channel diameter and single-stranded DNA
        persistence length, $D/\lss = 2$.}
\label{fig:g_m}
\end{figure}

The above expressions for $g(m)$ for a channel allow us to obtain numerical
results for the melting behavior. To that end we use the Fixman-Freire
approximation with stability parameters from Ref. \onlinecite{FK}, see
appendix \ref{sec:FF} for details. In practice only the first term of the sums 
in Eqs. (\ref{eq:gm_confined}) - (\ref{eq:GDs_resummed}) ($k=1$ 
and $n=1$) are sufficient (we denote them by $G_{\rm ss}^{(2)}$, $G_{\rm 
ds}^{(2)}$, $G_{\rm ss}^{(1)}$ and $G_{\rm ds}^{(1)}$ in the following).
We use $g_c(m)=G_{\rm ss}^{(1)}/G_{\rm ds}^{(1)}$ for small bubbles with size
$m<m_0$, $g_c(m)=G_{\rm ss}^{(1)}/G_{\rm ds}^{(2)}$ for the intermediate range
with bubbles of size $m_0<m<m_1$ and $g_c(m)=G_{\rm ss}^{(2)}/G_{\rm
  ds}^{(2)}$ for large bubbles with size $m>m_1$. The transition points $m_0$
and $m_1$ are found numerically as $m_0=0.7(D/\lss)^2$ and
$m_1=17.2(D/\lss)^2$ with maximum errors $0.016\%$ and $0.004\%$.
 Also, we ``pull out'' the first term in Eq. (\ref{eq:Gss}) and in
  Eq. (\ref{eq:GDs}) and redefine the hydrogen bond energies accordingly (see
  subsection \ref{sec:loop_factor_unconfined}), i.e., we write
\begin{equation}
g_c(m)= f(m) \exp \lpa \Big{(} \frac{2\Lambda^{\rm ds}_1}{x\lds}\kappa(\lds/D)
-\frac{4\Lambda^{\rm ss}_1 }{x\lss}  \Big{)} m \rpa
\label{eq:g_m_c1}
\end{equation}
Using the explicit form of the eigenvalues of the diffusion operator, we
have
\begin{equation}
g_c(m)= f(m) \exp\lpa\frac{\pi^2}{3xD^2}\lpa \lds\kappa(\lds/D)-2\lss\rpa m
\rpa\label{eq:g_m_c2}
\end{equation}
where we introduced a shifted loop factor $f(m)$. Introducing $f(m)$ and the
associated redefinition of the hydrogen bond energies (compare to subsection
\ref{sec:loop_factor_unconfined}), allows us to avoid numerical problems with
the FF algorithm due to multiplication of exponentially small numbers. The
function $f(m)$ is plotted in Fig. \ref{fig:g_m}, where we see that for small
bubbles we have $f(m)\propto m^{-3/2}$, whereas for large $m$ we have
$f(m)\propto m^{-1/2}$. The full functional form smoothly interpolates between
these two extremes. In Ref. \onlinecite{werner_schad2015} we show that $f(m)$
has a simple interpretation as being proportional to the probability that a
confined polymer forms a loop.

In the derivation of the loop factor $g(m)$ we make a number of simplifying 
assumptions. In particular, the ideal chain approximation is not a very 
realistic description of single stranded DNA. 
However, in
Ref.~\onlinecite{werner_schad2015} we find in simulations of a simple model of
DNA that the effect of confinement on the melting transition is qualitatively
similar for ideal and self-avoiding chains. For this reason, we expect that the
$g(m)$ used herein provides qualitatively correct predictions for the effect of 
channel confinement on the melting transition of DNA.

\section{Results}

The functional form for the loop factor $g(m)$ presented in the
previous section allows us to study how the melting transition changes
as the channel diameter is varied. We limit ourself to long DNA
molecules (thermodynamic limit) and study melting of three types of
prototypical DNA: 1. Homo-DNA, i.e. all basepairs are identical. For
this case we obtain analytical estimates alongside the numerical
results. 2. Random DNA, i.e., the A, T, G or C basepairs are chosen
with probabilities, $p_{A/T}$ and $p_{G/C}=1-p_{A/T}$. 3. Finally, we
consider a ``real'' DNA sequence, namely that of T4 phage DNA. The
numerical results are obtained by solving the Poland-Scheraga model
(Fixman-Freire approximation) using the functional form for $g(m)$ given
above. Details are found in appendix \ref{sec:numerics}.

\subsection{homo-DNA}\label{sec:melt_homo}

\begin{figure*}
  \begin{center}
    \includegraphics[width=17cm]{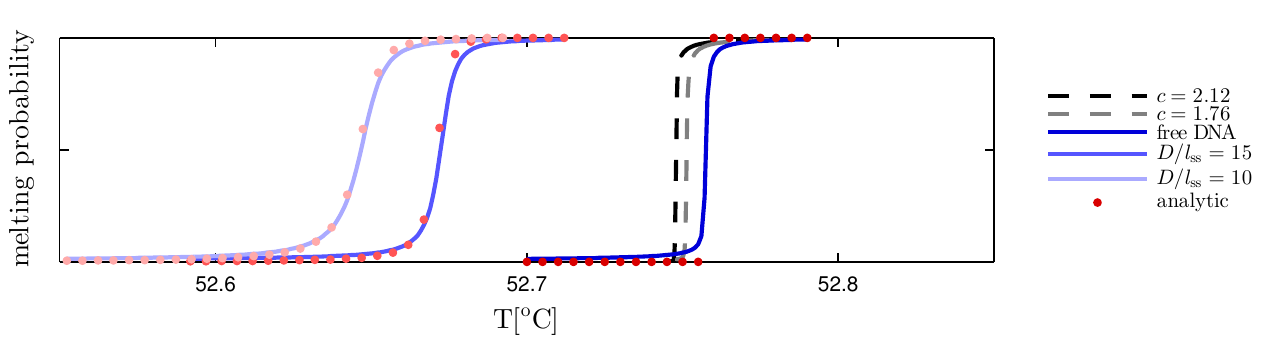}
  \end{center}
  \caption{ Melting probability as a function of temperature for homo DNA at
    different channel diameters. Numerical results (solid lines) and the
    analytical prediction (red marks) from Eq. (\ref{eq:P_1d}). Notice that
    decreasing the channel diameter leads to a decrease in the melting
    temperature and to a broadening of the melting curves.  The analytical
    prediction agrees rather well with the numerical results already for
    moderate confinement. Dashed lines show melting probabilities for
    different values of the loop exponent $c$. We see only a very slight
    change of the width of the transition, for further details on the effect
    of $c$ and comparison to the analytic prediction see
    figs. \ref{fig:width_vs_c} and \ref{fig:melting_homo_c} in the
    appendix. Parameters: number of basepairs = $2\cdot 10^5$, input sequence
    = `AAAAA ... AAAA', and salt concentration = 0.01 $M$. Note that 'free
    DNA' here refers to the case $c=1.5$ (unconfined ideal polymer).}
\label{fig:melt_homo}
\end{figure*}

We first consider homo-DNA, i.e. assume that the parameters $E_{\rm
  hb}(i)$ and $E_{\rm st}(i-1,i)$, see subsection \ref{sec:general}, are
independent on $i$. We first present analytical estimates for the
melting curve, the melting temperature, and the width of the melting
curve for small channels, before comparing to numerical results.

In the limit of small channels (or large bubbles), Eq. (\ref{eq:gm_confined}) 
becomes 
\begin{align}\label{eq:gm_large}
\lim\limits_{m \to \infty}g_c(m) &=\frac{G_{\rm ss}^{(2)}}{G_{\rm 
ds}^{(2)}} = \frac{\pi^4}{64} \lpa\frac{2m}{x\lss}\rpa^{-1/2}\nonumber \\
& \times \!A 
 \exp\lpa\frac{\pi^2}{3xD^2}\lpa \lds\kappa(\lds/D)-2\lss\rpa m\rpa
\end{align}
Using this approximation, we can derive a number of exact 
results.

\begin{figure}
  \begin{center}
    \includegraphics[width=8.5cm]{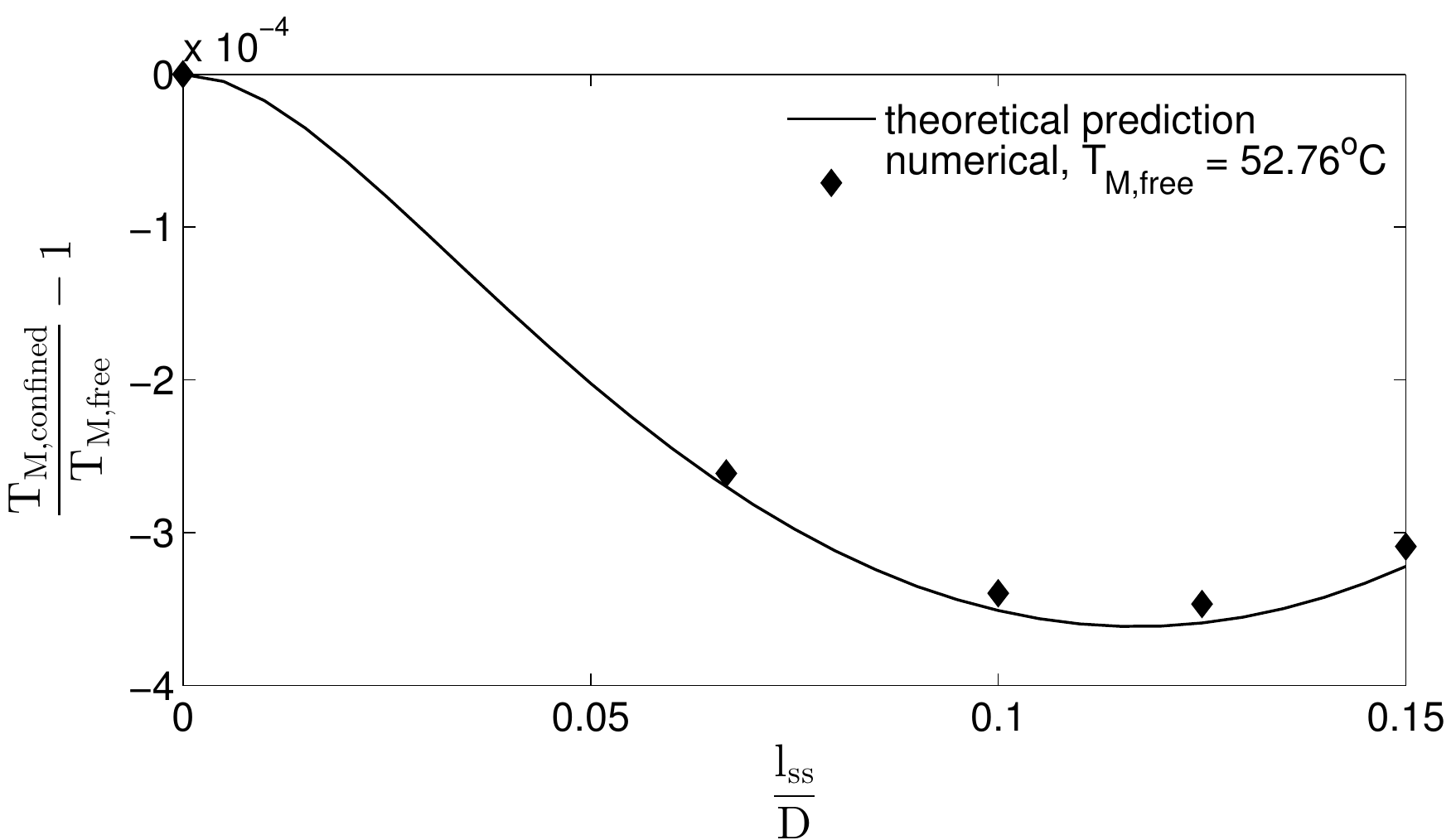} \\
    \hspace*{0.5cm}\includegraphics[width=8cm]{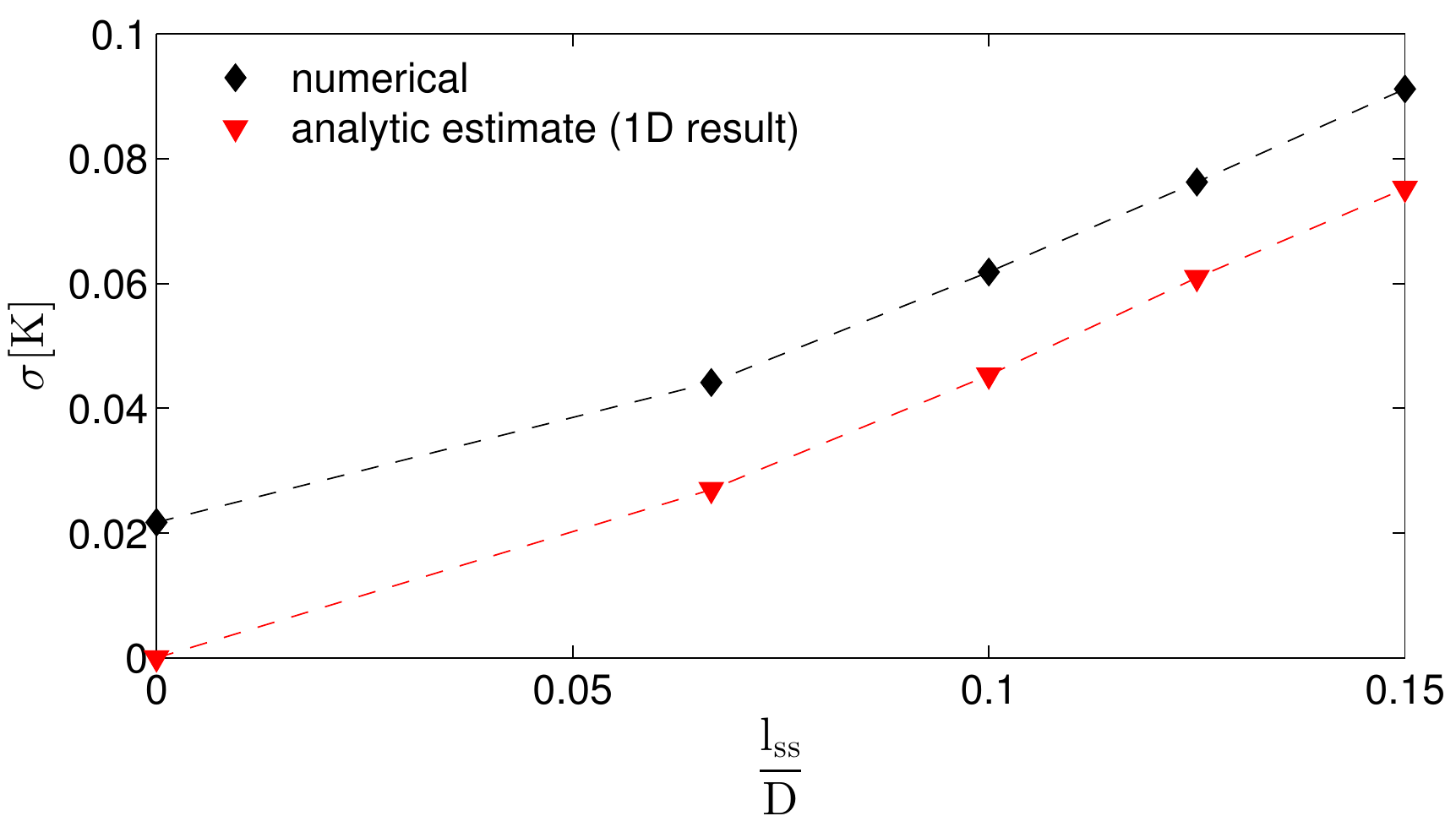}
  \end{center}
  \caption{Confinement
    induced shift in melting temperature (units of  Kelvin) and width of 
melting 
curves, for
    homo-DNA, as a function of inverse channel diameter. (Top) Shift of the
    melting temperature, where the theoretical prediction is given in Eqs.
    (\ref{eq:TMshift}) and (\ref{eq:DeltaS}). (Bottom) Width of the melting 
curve, where the
    theoretical result is obtained through Eq. (\ref{eq:P_1d}), see main text
    for details. Input parameters were the same as in
    Fig. \ref{fig:melt_homo}. }
\label{fig:TM_homo}
\end{figure}

First, there is a channel induced shift in the melting temperature: as
discussed in the previous section, the exponential term in Eq.~(\ref{eq:g_m_c2}) 
can be included in
the hydrogen bond energies, $E_{\rm hb} \rightarrow E_{\rm hb} -
RT\frac{\pi^2}{3xD^2}\lpa \lds\kappa(\lds/D)-2\lss\rpa $ in analogy with how we 
included the coordination
number in Sec. \ref{sec:loop_factor_unconfined}. This leads to a shift of the
melting temperature according to
\be\label{eq:TMshift} 
T_{\rm{M,confined}}=\frac{U}{S}=\frac{S_{\rm{ref}}}{S_{\rm{ref}}-\Delta S 
}T_{\rm M} \ee 
with  
\be\label{eq:DeltaS}
 \Delta S = \Delta S_{\rm ss} -\Delta S_{\rm ds} = R \frac{\pi^2}{3xD^2}\lpa 
\lds\kappa(\lds/D)-2\lss\rpa 
\ee 
and where $R$ is the molar gas constant.  Confining an ideal (linear) chain to
a channel decreases the entropy by an amount $\Delta S_{\rm ss}$ per link for
the single-stranded (melted) region and $\Delta S_{\rm ds}$ for the intact
double stranded region. This entropy of confinement agrees with a scaling
argument given in section I.1.3 in Ref. \onlinecite{degennes}. For
$\lds\kappa(\lds/D) > 2\lss$ we predict that there is an entropy driven
decrease in melting temperature for small channels. In particular, we show
in Fig. \ref{fig:TM_homo} that for typical experimental channel sizes,
confinement indeed decreases the melting temperature of DNA. Moreover, we find 
that the melting temperature as a
function of $\lss/D$ has a minimum at around $\lss/D\approx 0.12$. Within
our model, this finding follows from the fact that a double-stranded region
enters the Odijk regime (where the entropy of confinement scales as
$(\lds/D)^{2/3}$ per Kuhn length [see Eqs. 
(\ref{eq:GDs}-\ref{eq:kappaEquation})], rather than as
$(\lds/D)^2$ as for the Gaussian chain regime) at a larger value of $D$
than a single-stranded region. Note that we for the single-stranded region
actually do not include the Odijk regime in our expression for the
associated Green's function. Hence, we
cannot realistically consider larger values for $\lss/D$ than those shown
in Fig. \ref{fig:TM_homo}.  We here note that in the simplified model in
Ref. \onlinecite{werner_schad2015} the persistence length of single-stranded
and double-stranded regions are the same which rather leads to a slight
increase of the melting temperature for decreasing channel diameters.

Second, once we have redefined the melting temperature, we can in fact predict
the full melting curve. To that end we define an effective ring factor as
$\xi_{\rm eff} = (\pi^4/64) (x\lss/2)^{1/2} A\xi = K (\lss/D)^2\xi $, with $K=
\pi^5/(48 x\lss)$; for $\lss$= 6 nm we have $K = 0.354$. Then the problem at
hand becomes that of melting with a loop factor $\sim m^{-1/2}$
(``one dimensional'' random walk) and ring factor $\xi_{\rm eff}$ (all valid
for small channels). The change in the value of the ring-factor,
$\xi\rightarrow \xi_{\rm eff}$, is again due to entropic confinement effects.
We here restrict ourselves to $D$ values such that $D/\lss \ge 7$, for which
we have $\xi_{\rm eff}<\xi$. For smaller values of values of $D/\lss$ the
diffusion approximation used for the single-stranded regions breaks down, see
previous section.  Using standard analytical techniques (appendix
\ref{sec:homo_DNA}) we find that the expected fraction of melted basepairs is
given by:
\be\label{eq:P_1d} P =1-\frac{1}{1+\sigma_0 \beta^{-1}z_0 {\rm
    Li}_{c-1}(z_0)} 
\ee 
where $c=c_{\rm 1d} = 1/2$ and $\beta=\exp[\Delta S (T-T_{\rm M,confined})/(R
T)]$ with the ``universal'' dimensionless DNA constant $\Delta S/k_B\approx
12.51$ [since $R=1.987$ cal/(mol K) and the entropy loss is $\Delta S=24.85$
cal/(mol K), see Ref. \onlinecite{FK}]. The quantity $\sigma_0 = \xi_{\rm eff}
\exp [-E_{\rm st}/(GT)]$ is the cooperativity parameter, here modified to take
into account effects due to the channel (through $\xi_{\rm eff}$). The
quantity $z_0$ is obtained by solving Eq. (\ref{eq:z0}) numerically; whenever
no root $z_0$ is found in the interval $0 < z_0 < 1$, one sets $P = 1$.

Let us now present results of numerical simulations. In
Fig. \ref{fig:melt_homo} we display the melting probability for different
channel diameters as a function of temperature. The numerical results are 
compared to the analytical prediction of Eq. (\ref{eq:P_1d}). We also include 
results
for unconfined DNA with $c=1.76$ and $c=2.12$. The homo-DNA melting curves are
generated by using as input a sequence $AAA ... AAA$ to our Fixman-Freire
code.  We find that as the channel diameter is decreased the melting
temperature increases and the width of the transition increases. The melting
curve is slightly asymmetric in agreement with the prediction in
Eq. (\ref{eq:P_1d}). However, the asymmetry is less pronounced for melting in
the channel compared to melting of unconfined DNA ($c=3/2$, $c=1.76$ and
$c=2.12$). In Fig. \ref{fig:melt_homo} we also show the analytical prediction
from Eq. (\ref{eq:P_1d}) and find good agreement with the numerical
results. As input for the unconfined DNA melting temperature, $T_M$,
  appearing in Eq. (\ref{eq:TMshift}), we use the temperature at which $P=1/2$
  from simulations for the unconfined case $(c=3/2$). Note that $T_M$ obtained
  in this way deviates by roughly 0.03 $^\circ$C degrees compared 
to the
  Marmur-Doty formula. 

Figure \ref{fig:TM_homo} shows results for the melting temperature (top) and
width (bottom) of the melting curve as a function of channel diameter. These
results where extracted from melting curves like those presented in Figure
\ref{fig:melt_homo}: the melting temperature is the temperature at which we
have $P=1/2$ and the width $\sigma$ of the melting transition is obtained as
$\sigma=T(f=0.97)-T(f=0.03)$. We find excellent agreement with the analytical
prediction [Eqs. (\ref{eq:TMshift}) and (\ref{eq:DeltaS})] for the decrease in 
melting temperature
with decrease in channel diameter.  We get our analytic prediction for
$\sigma$ by setting $P=P_{\rm upper} = 0.97$ and $P=P_{\rm lower}=0.03$ in
Eq. (\ref{eq:P_1d}) and then solve numerically (bisection method) in order to
obtain temperatures $T_{\rm upper}$ and $T_{\rm lower}$ respectively. From
these solutions we then calculate $\sigma = T_{\rm upper}-T_{\rm lower}$. Our
results show that stronger confinement leads to an increase in the width of
the melting transition. At strong confinement we 
find good agreement between the numerical results and our analytical prediction.
Since we assume $\lss$ = 6nm, the choice of nanochannel
diameters considered in Fig. \ref{fig:TM_homo} corresponds to 40 nm$\ \le D
\le \ $90 nm. 

\subsection{Random DNA}

\begin{figure*}
  \begin{center}    
\includegraphics[width=17cm]{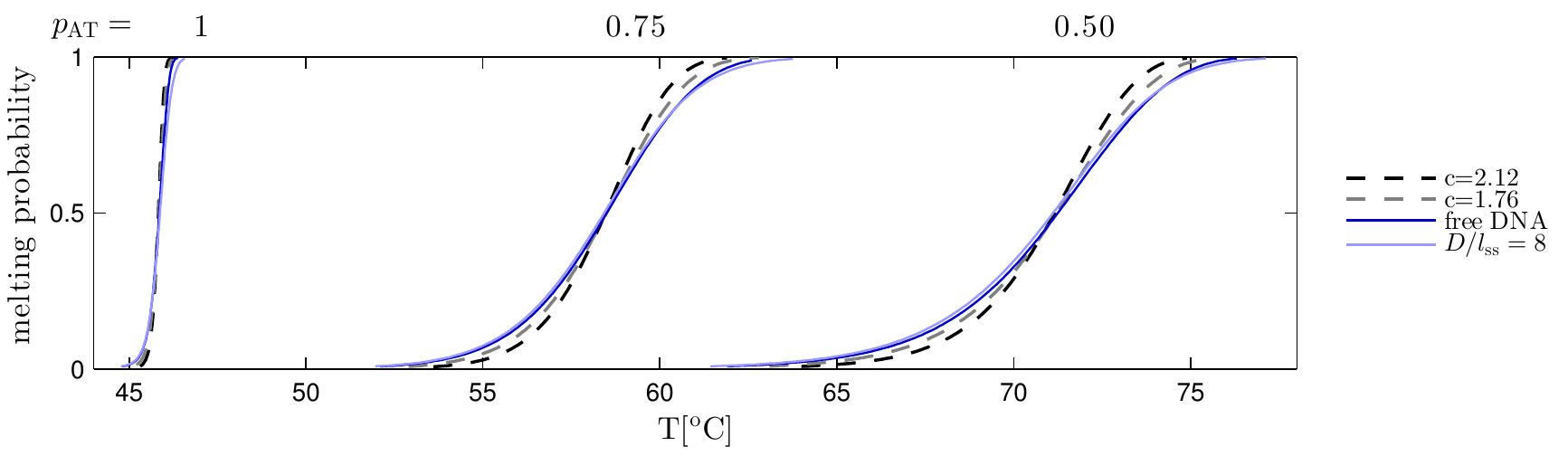}
  \end{center}
  \caption{ Average
    melting probability for random DNA (with different AT fractions) as a
    function of temperature at different channel diameters. For a fixed AT
    fraction, $p_{\rm AT}$, we find that smaller channel diameters leads to
    decreased melting temperatures and a broader transition (compare to
    Fig. \ref{fig:melt_homo}). The main effect of the heterogeneity (non-zero
    $p_{\rm AT}$) is to further broaden the transition compared to melting of
    homo-DNA -- the transition is broadest for the case $p_{\rm
      AT}=1/2$. Parameters: number of basepairs = $2\cdot 10^5$, and salt
    concentration = 0.01 $M$.  Each curve is an average over 200 random
    sequences of length 200 kilo basepairs with a given AT-fraction, $p_{\rm
      AT}$. Note that `free DNA' here refers to the case $c=1.5$ (unconfined
    ideal polymer).  }
\label{fig:melt_random}
\end{figure*}

In order to investigate how the sequence affects the melting transition, in
this section we present numerical results for confined random DNA (for such a
scenario no analytical predictions are available). In all simulations we
generated 200 random sequences with a given AT-fraction, $p_{\rm AT}$, then
used the Fixman-Freire approximation to predict a set of probability
profiles. Finally, we calculated the average melting probability from this
set.

\begin{figure}
  \begin{center}
    \includegraphics[width=8.5cm]{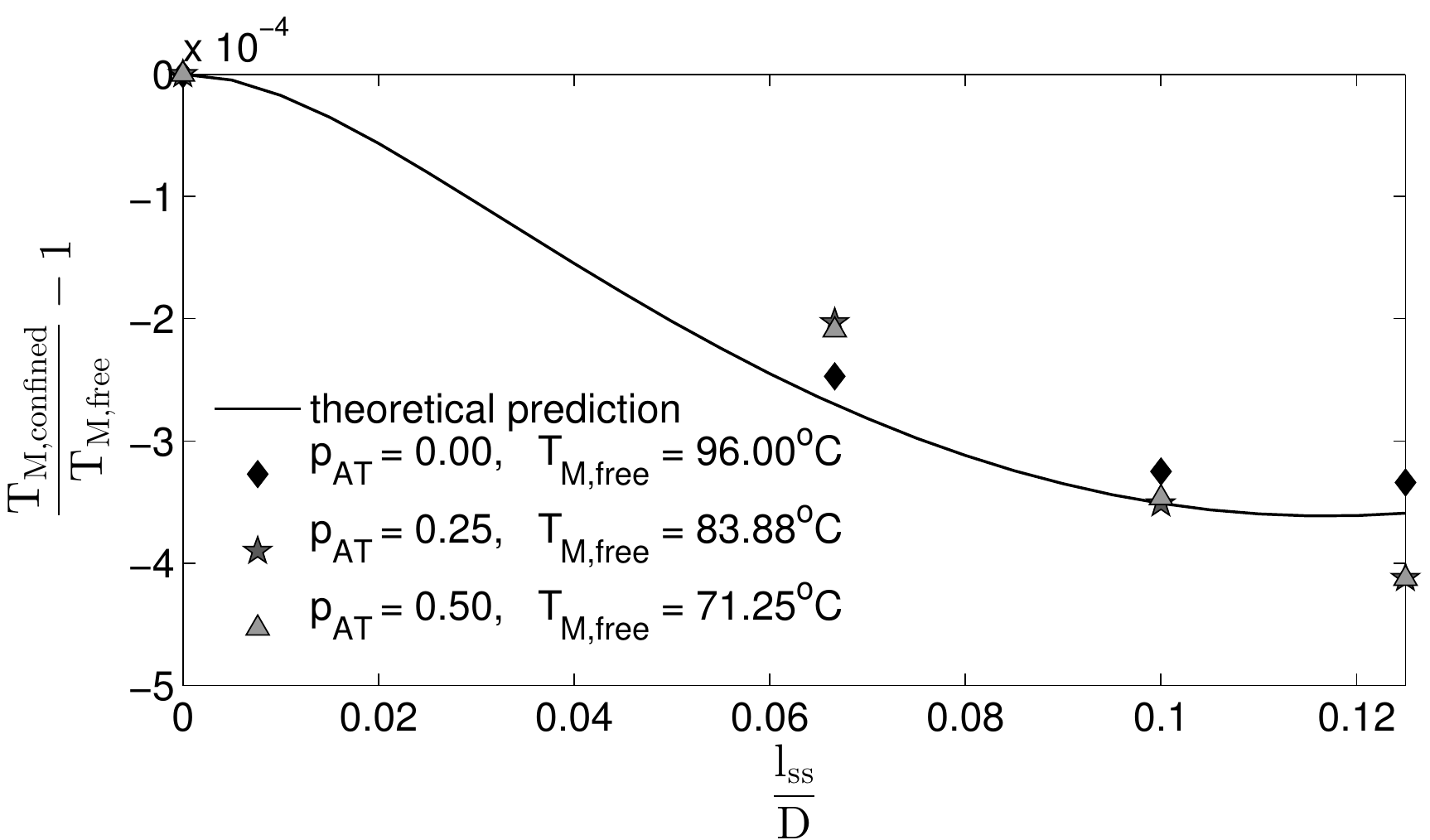} \\
    \hspace*{0.5cm}\includegraphics[width=8cm]{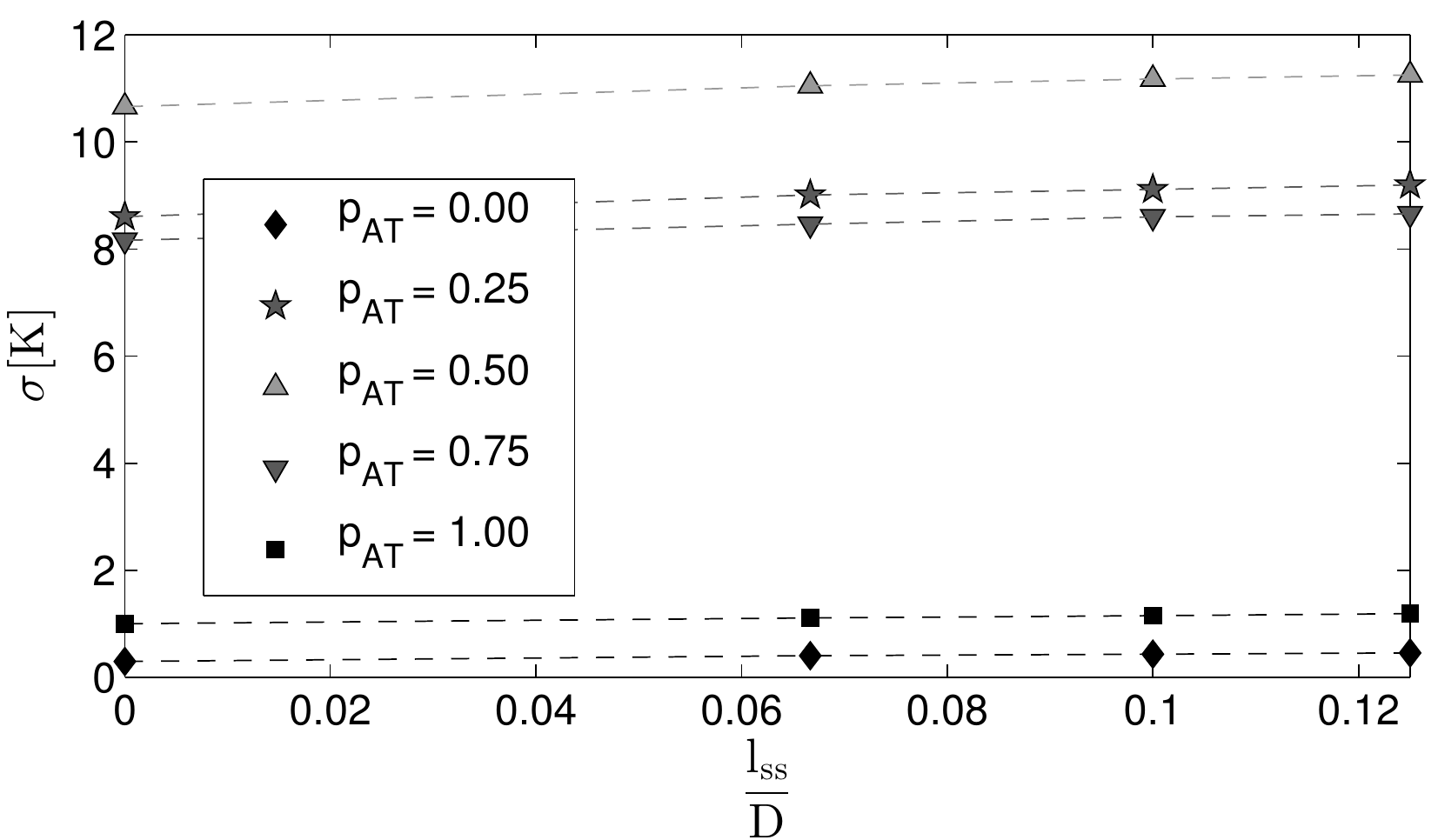}
  \end{center}
  \caption{Confinement induced shift in melting temperature  (units of
    Kelvin) and width
    of melting curves, for random DNA with AT fraction $p_{\rm AT}$ ,
    as a function of inverse channel diameter.  (Top) Shift of the melting
    temperature. Interestingly, the shift in melting temperature,
    Eqs. (\ref{eq:TMshift}) and (\ref{eq:DeltaS}), obtained for a homo-DNA 
scenario, works
    rather well also for random DNA.  (Bottom) Width of the melting
    curve. We notice that the width increases with decreasing channel
    diameter, and that the width is at maximum for $p_{\rm AT}=1/2$.
    Input parameters were the same as in Fig. \ref{fig:melt_homo}.
    }
\label{fig:TM_random}
\end{figure}

In Fig. \ref{fig:melt_random} we display the average melting
probability for different channel diameters as a function of
temperature for three different $p_{\rm AT}=\{0,0.25,0.5\}$. The results 
from Fig. \ref{fig:melt_random} are further analyzed in 
Figure \ref{fig:TM_random} which shows results for the melting temperature
(top) and width (bottom) of the melting curve as a function of channel
diameter. These results where obtained from melting curves as
displayed in Fig. \ref{fig:melt_random} in an identical fashion as in the
previous subsection.

Figs.  \ref{fig:melt_random} and \ref{fig:TM_random} show a number of 
 interesting results. The main effect of the
heterogeneity (non-zero $p_{\rm AT}$) is to further broaden the
transition -- the transition is broadest for the case $f_{\rm
  AT}=1/2$, and sharpest for homo-DNA, $p_{\rm AT}=1$ (pure AT)
and $p_{\rm AT}=0$ (pure GC). Just as for the homo-cases (see previous
subsection) we find, for a fixed value of $p_{\rm AT}$, that
increasing confinement leads to a decreased melting temperature and
an increase in the width of the transition.
Interestingly, the shift in melting temperature,
Eqs. (\ref{eq:TMshift}) and (\ref{eq:DeltaS}), obtained for a homo-DNA scenario, 
works
rather well also for random DNA, see Fig. \ref{fig:TM_random} (top). 

Studying Fig. \ref{fig:TM_random} (Bottom) we see that while confinement 
broadens the melting transition also for random DNA, the effect is relatively 
small 
compared to the intrinsic width of the transition, which is much larger for 
random 
DNA compared to the homo-DNA case. This effect is also seen in Fig. 
\ref{fig:melt_random}, where we find
that the difference between ``free DNA'' and confined random DNA is small.

\subsection{T4 phage DNA molecules}

We now investigate a ``real'' DNA sequence - T4 phage, and compare to the 
melting
behavior for random DNA.

\begin{figure}
  \begin{center}
  \includegraphics[width=8.8cm]{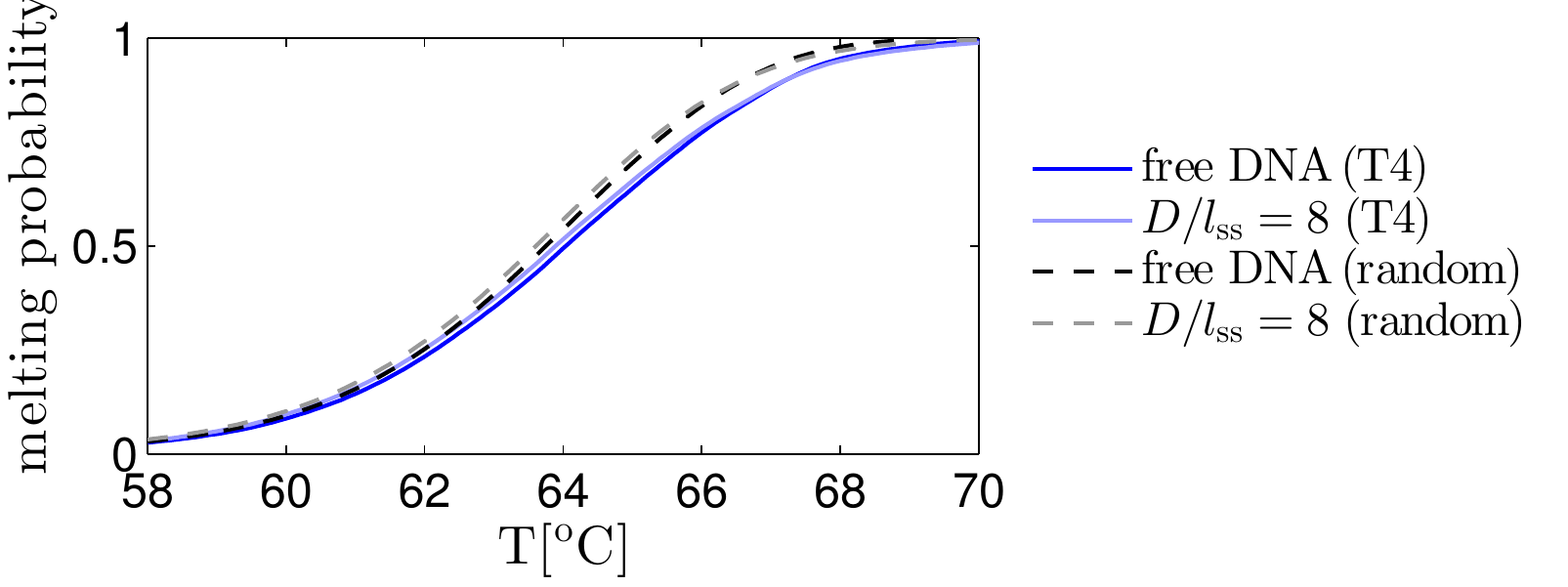}\\
  \end{center}
  \caption{Melting probability for a T4 phage DNA, and random DNA, as
    a function of temperature at different channel diameters. The
    melting behavior of T4 phage DNA (same $p_{\rm AT}$ and number of 
basepairs as the T4 phage DNA sequence) is very similar to its
    associated random sequence melting curve. However, the random DNA
    melting curve is somewhat steeper. Parameters: number of basepairs
    = $168900$, and salt concentration = 0.01 $M$.  Each random
    curve is an average over 200 random sequences with a given
    AT-fraction, $p_{\rm AT}=0.647016$. The sequence for T4 phage was 
downloaded from the NCBI GenBank (NC\_000866.4).}
\label{fig:melting_T4}
\end{figure}

\begin{figure}
  \begin{center}
    \includegraphics[width=8.5cm]{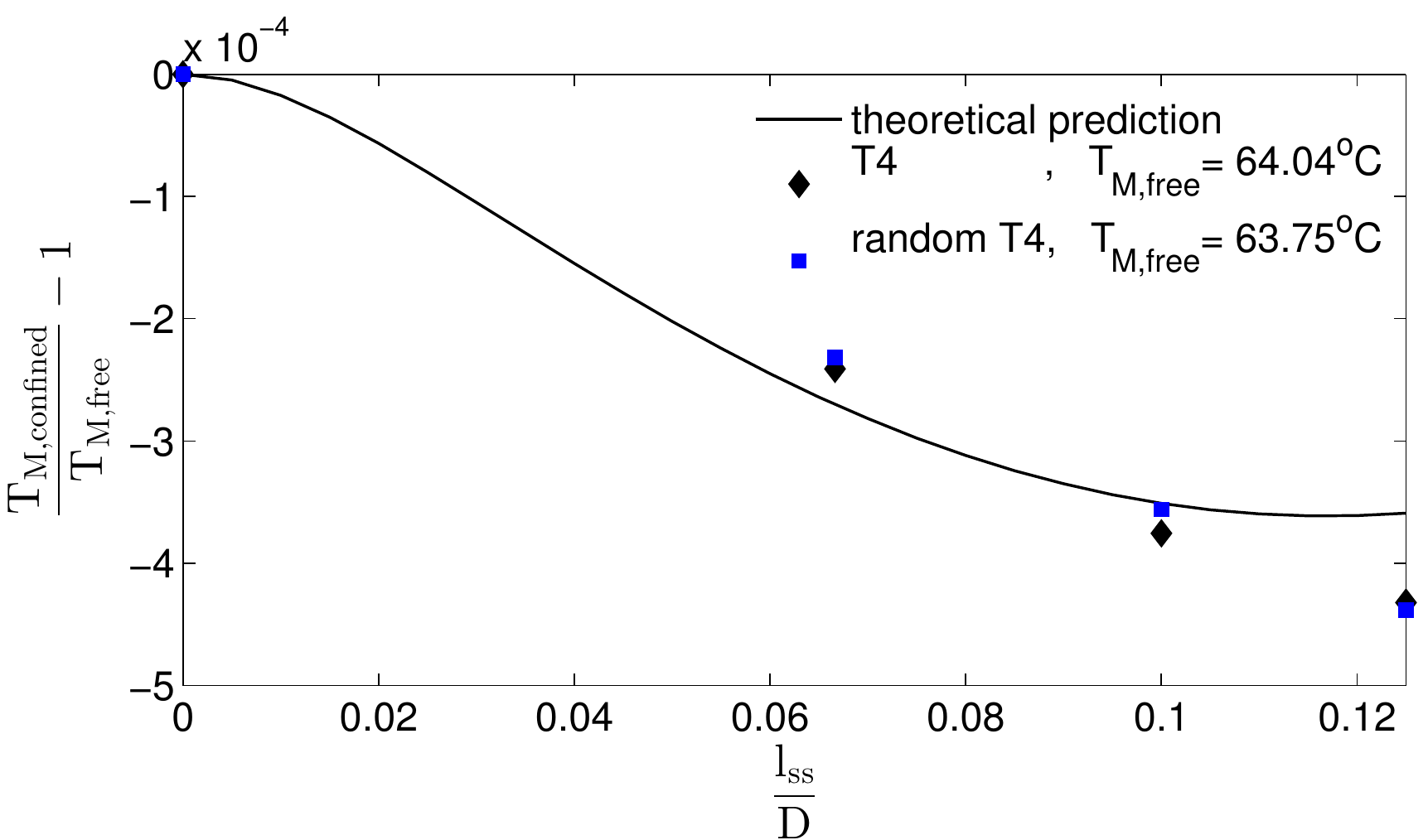}\\
    \hspace*{0.5cm}\includegraphics[width=8cm]{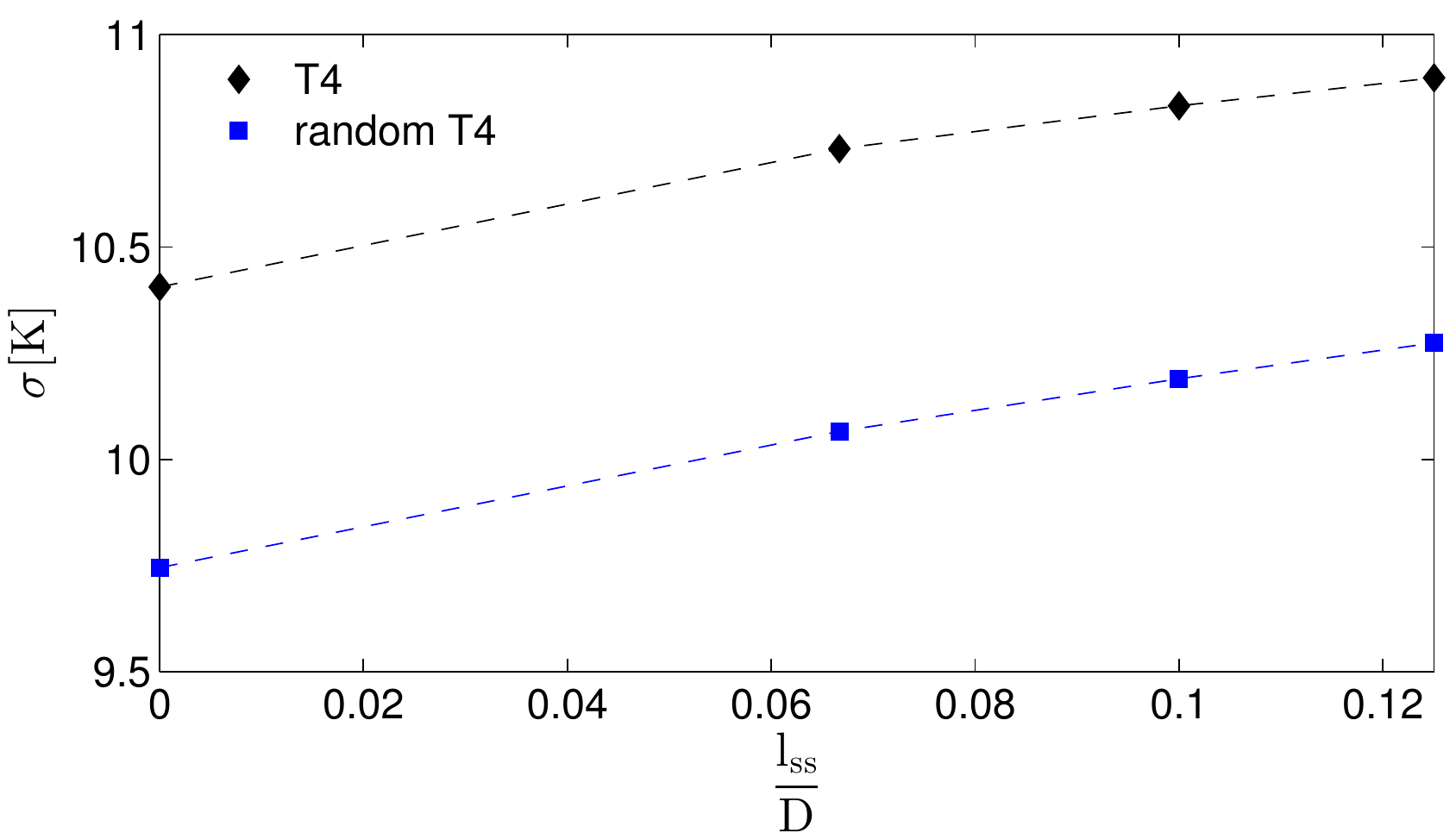}\\
  \end{center}
  \caption{Shift in melting temperature  and width of melting curves T4 phage
    DNA and associated random DNA (same AT fraction), see
    Fig. \ref{fig:melting_T4}, as a function of inverse channel diameter. (Top)
    Change of the melting temperature  (units of  Kelvin)due to confinement, 
where the solid
    curve is Eqs. (\ref{eq:TMshift}) and (\ref{eq:DeltaS}). Notice that the 
shift in melting
    temperature for T4 phage DNA is very similar to the melting temperature
    for a random DNA with the same AT fraction. (Bottom) Width of the melting
    curve at different confinement. The width of the melting curve for T4
    phage DNA is slightly larger than the width for a random DNA with the
    same AT fraction. Input parameters were the same as in
    Fig. \ref{fig:melting_T4}. }
\label{fig:TM_T4_D}
\end{figure}

In Fig. \ref{fig:melting_T4} we display the melting probability for free DNA
and DNA confined to a channel with different diameters as a function of
temperature for T4 phage and for random DNA with the same length
$L=165643$ and AT ratio $p_{\rm{AT}}=0.64711$. Figures \ref{fig:TM_T4_D} (top, 
bottom) show the shift of the
melting temperature and the change of the width $\sigma$ for different  
confinement.

From Figures \ref{fig:melting_T4} and \ref{fig:TM_T4_D} a few things
are worth pointing out. First, we notice that the melting curves for
random DNA and T4 phage DNA are very similar. The main difference is
that the width is somewhat increased for T4 DNA (however, less than 10
$\%$ for all channel diameters investigated). Obviously, these
findings cannot be generalized to arbitrary ``real'' sequences, but
shall, rather, be used as a motivation for why studying random DNA may be
relevant also for DNA melting experiments.

\section{Summary and outlook}

In this study we investigate how confinement affects the DNA melting
transition. The statistical physics description of this process has a long 
history, and is commonly described in terms of the Poland-Scheraga (PS) model. 
The PS model is an Ising type model with heterogeneous interaction due to the 
sequence dependent energies required for breaking the bonds between the 
complementary bases. Moreover, including the configurational entropy associated 
with a melted DNA region gives rise to an effective long range interaction. It 
has been found previously that this loop entropy factor changes the nature of 
the DNA melting transition -- from ``smooth'' in the absence of this factor, to 
first order when the loop factor is included.\cite{kafri} These results pertain 
to a scenario of identical basepairs along an unconfined DNA.

Inspired by recent experiments,\cite{reisner2010,reisner2012} we calculate the
effect that confinement in a nanochannel has on the melting transition of DNA.
By subsequent numerical solutions of the PS model, and simplified arguments,
we demonstrate that the confinement has three effects: the melting
temperature is decreased, the confinement broadens the melting transition, and
the shape of the melting curve goes from a functional form for unconfined
melting (loop exponent $c=3/2$) to a form consistent with one-dimensional
melting ($c=1/2$). We further quantify the sequence dependence of the
melting transition, and find that the transition is broadest when the fraction
of AT basepairs is 1/2. Relative to the width of the transition in the 
unconfined case, the broadening effect of confinement is smaller for random DNA 
than for homo-DNA. We also investigate ``real'' DNA, here T4 phage DNA,
and find that its melting behavior is very similar to the melting of a random
DNA with the same AT fraction.  This finding gives further importance
to recent efforts for understanding DNA melting of unconfined random 
DNA.\cite{hwa,garel_04,giacomin2009,alexander2009,derrida2014}

Throughout this study we use a functional form of the loop factor
$g(m)$ for a confined chain, which is derived assuming ideal
chain statistics (no self-exclusion
effects included), and relaxing the constraint that within a melted region no
complementary base pair binding must take place.\cite{werner_schad2015}
While these approximations imply that one cannot expect the predictions of 
this study to be exactly valid for real DNA, we find a qualitatively similar 
effect of confinement upon the melting transition for simulations of an ideal 
DNA model and of a self-avoiding one in Ref.~\onlinecite{werner_schad2015}. We 
therefore expect that the results of this study furnish a qualitatively correct 
description of the effect of confinement on the melting transition of real DNA.

In a broader perspective, our study sheds light on the DNA strand separation
in cells. The cell interior is a confined and crowded environment, where the
crowding due to other macromolecules is known to affect vital biological
processes such as the folding of proteins, diffusive transport and reaction
rates.\citep{Ellis_2003} Our study serves as a first step towards an
understanding of how confinement and crowding affects the DNA melting, with
potential implications for understanding of the processes whereby proteins 
access single strands of DNA,\cite{mark,mark1} and for transcription 
initiation\cite{kalo,yeramian} in living cells.

\section{Acknowledgment} 

TA is grateful to the Swedish Research Council for funding (grant no
2009-2924 and 2014-4305). BM acknowledges financial support
by the Swedish Research Council and by the G\"oran Gustafsson
Foundation for Research in Natural Sciences and
Medicine.

\appendix

\section{Numerical implementation}\label{sec:numerics}

In this appendix we introduce one exact scheme, the Poland algorithm, and one
approximate scheme, the Fixman-Freire approximation, for computing DNA melting
profiles.

Let us first introduce some notation. The full sequence of basepairs enters
via the position-dependence of the statistical weights
\be\label{eq:alpha}
\alpha_i=\exp\{-E_{\rm hb}(i)/[RT]\}
\ee
 for breaking the hydrogen-bonds of the basepair at position $i$, and
\be\label{eq:delta}
\delta_i=\exp\{-E_{\rm st}(i-1,i)/ [RT]\}
\ee
for disrupting the stacking interactions between basepairs $i-1$ and $i$. In
addition, we need the parameters $\xi$ and the loop factor $g(m)$, see main 
text.

\subsection{The Poland algorithm}

We now provide an exact scheme, the Poland algorithm\cite{poland_algo} for 
clamped ends, for computing the probability that basepair $k$
is open. The algorithm presented here extends previous methods by incorporating
explicit stacking interactions: previous approaches ``lumped'' one Boltzmann
factor of stacking interactions together with the $\xi$, giving rise to a
cooperativity parameters, $\sigma_0$. 

The Poland algorithm has two components. First, the conditional probability
$P_c(k)$ that basepair $k+1$ is closed, provided that basepair $k$ is closed, is
computed. Note that we here use the (somewhat funny) short-hand
notation $P_c(k)=P_c(k+1|k)$ following Poland.\cite{poland_algo} 
Secondly, the unconditional probability $P_u(k)$ that basepair
$k$ is closed, is computed. In practice, however, one does not directly work 
with $P_c(k)$ but rather introduces $T_k=\alpha_k\delta_kP_c(k)$. If we now 
extend the algorithm from Ref. \onlinecite{poland_algo} to the case of clamped 
boundary conditions and explicit stacking parameters, we find the following 
recursion relation:
\bea\label{eq:Poland_short1}
T_N&=&\alpha_N\delta_N,\nonumber\\
T_k&=&\alpha_k\delta_k [1+\sum_{j=1}^{N-k}a_k(j)]^{-1}, \ 
k=N-1,N-2,...,0,\nonumber\\
\eea
where $a_k(j)\equiv q_I(k,j)\prod_{l=k+1}^{j+k}P_c(l)$ with the statistical 
weight $q_I(k,j)$ for a bubble with its left boundary
at position $k$ (basepair $k$ is closed and basepair $k+1$ is open) and 
consisting of $j$
consecutive open basepairs:
\be\label{eq:q_I}
q_I(k,j)=\xi g(j)\prod_{l=k+1}^{j+k} \alpha_l \prod_{l=k+1}^{j+k+1} \delta_l.
\ee
We above defined
$\alpha_0=\delta_0=1$. The quantity $a_k(j)$ is conveniently computed using
the recursion relation
\bea\label{eq:Poland_short2}
a_k(1)&=&\xi g(1)\delta_{k+2}T_{k+1}\qquad k=0,1,...,N-1,\nonumber\\
a_k(j)&=&\frac{g(j)}{g(j-1)}T_{k+1}a_{k+1}(j-1)\qquad j=2,...,N-k.\nonumber\\
\eea
The (unconditional) probability $P_u(k)$ that basepair $k$ is then obtained from 
the 
equations:
\bea\label{eq:Poland_short3}
P_u(0)&=&1,\nonumber\\
P_u(1)&=&T_0,\nonumber\\
P_u(k+1)&=&P_u(k)P_c(k)+\sum_{j=0}^{k-1}\mu_k(j)P_u(j),\ k=1,2,...,N.\nonumber\\
\eea
We further define $\mu_k(j)=q_I(j,k-j)\prod_{l=j}^k P_c(l)$ from which we find 
the
recursion relation:
\bea\label{eq:Poland_short4}
\mu_k(k-1)&=&\xi g(1)
\frac{\delta_{k+1}}{\alpha_{k-1}\delta_{k-1}}T_{k-1}T_k\qquad k=2,3,...,N, 
\nonumber\\
\mu_k(j)&=&\frac{g(k-j)}{g(k-j-1)}\frac{\delta_{k+1}}{\delta_k}T_k
\mu_{k-1}(j)\nonumber\\
&& j=0,1,...,k-2.
\eea
The Poland algorithm for clamped DNA and explicit stacking parameters thus 
becomes: First use Eqs. (\ref{eq:Poland_short1}) and (\ref{eq:Poland_short2}) 
to obtain $T_k$ for $k=N,N-1,...,0$. Then use Eqs. 
(\ref{eq:Poland_short3}) and (\ref{eq:Poland_short4}) to obtain $P_u(k)$ for 
$k=0,1,.....,N+1$. A consistency check of the calculation is that 
$P_u(N+1)\equiv 1$ due to the clamping.

The mean number of bubbles used in appendix \ref{sec:deviation_melting_temp} 
can easily be determined from $P_u(k)$ and $P_c(k)$ as the number of left 
boundaries (basepair $i$ closed, basepair $i+1$ open):
\bea\label{eq:N_B}
\langle N_B\rangle  &=& \sum_i \langle s_i (1-s_{i+1})\rangle \nonumber\\
&=& \sum_i P_u(i) - \sum_i P_u(i) P_c(i)
\eea
where we introduced the ``spin'' of a basepair: $s_i = 1$ if basepair i 
is closed and $s_i = 0$ otherwise.

The Poland algorithm scales with the number of basepairs $N$ as $N^2$. In 
practice, we therefore do not use the Poland algorithm for $N > 10$ kbp. 
In the next section we present an algorithm whic scales linear in $N$, the 
Fixman-Freire approximation.\cite{fixman}

\subsection{Fixman-Freire approximation}\label{sec:FF}

The Fixman-Freire (FF) approximation uses the fact that a sum of exponentials
is a good approximation for a power law function. The sums in
Eqs. (\ref{eq:Poland_short1}) and (\ref{eq:Poland_short3}) give rise to the 
$N^2$
scaling of the Poland algorithm. Replacing the loop factor $g(m)$ (as
part of both sums, see Eq. (\ref{eq:q_I})) by a sum of exponentials with $I$
terms, reduces the scaling of the approximation to an $N I$-scaling.

Thus, within Fixman-Freire's method, the following approximation is introduced:
\bea\label{eq:FF0}
g(m)\approx \sum_{i=1}^{I} c_i \exp(-b_i m)
\eea.
Applied on the sum in Eq. (\ref{eq:Poland_short1}), we find
\be\label{eq:FF1} 
T_k = \alpha_k\delta_k [1+ \sum_{i=1}^{I}c_i S_i(k)]^{-1}
\ee
and the recursion
relation
\bea\label{eq:FF2}
S_i(N) &=& 0 \nonumber\\
S_i(k-1) &=& T_k \exp(-b_i) [\xi \delta_{k+1} +S_i(k)].
\eea
Similarly, Eq. (\ref{eq:Poland_short3}) becomes
\be\label{eq:FF3}
P_u(k+1)=P_u(k)P_c(k)+\sum_{i=1}^{I} c_i A_i(k)
\ee
with recursion relation
\bea\label{eq:FF4}
A_i(0) &=& 0 \nonumber\\
A_i(k+1) &=& P_c(k+1) \alpha_{k+1} \delta_{k+1} \exp(-b_i) \nonumber\\
&&\left[ \xi \delta_{k+1} P_u(k) P_c(k) +A_i(k)\right].
\eea
The coefficients $c_i$ and $b_i$ are determined from a fit on $2I$ points, 
$l_i$ (chosen such that the log's are equally distributed on the domain where 
the approximation should be valid, i.e. from 1 to the total length of the DNA
sequences). Starting with all coefficients set to zero, first the guess values
$c_I$ and $b_I$ are determined from $g(l_{2I})$ and $g(l_{2I-1})$, then follow
$c_{I-1}$ and $b_{I-1}$ using the new set of parameters and so on. If no
convergence can be achieved, the parameters are determined for a larger
interval, which is then slowly shrunk again with the new set of parameters as
a first guess.

\section{DNA melting behavior for homo-DNA}\label{sec:homo_DNA}

In this appendix we analyze exact recursion relations for the partition
function  for the Poland-Scheraga model, and provide an analytic expression for
the melting probability for homo-DNA.

Following Ref. \onlinecite{garel_04} we divide the DNA molecule into a left
region, i.e. basepairs $\{1,...,k-1\}$ and a right region, basepairs 
$\{k+1,...,N\}$,
where $k$ varies. The end basepairs, basepairs number 0 and $N+1$, are here
assumed clamped (closed). Garel and Orland, see Ref. \onlinecite{garel_04},
proceed by deriving a recursion relation (see below) for the statistical
weights (partition functions), $Q_L(k)$ and $Q_R$(k), for the ``left'' and
``right'' regions with the constraint that basepair $k$ is {\em closed}. Given
$Q_L(k)$ and $Q_R(k)$ we can write most quantities of interest. In particular,
the probability for having basepairs $k$ closed is simply
\be\label{eq:P_open}
P (k) = 1 -\frac{Q_L(k)Q_R(k)}{Q_L(N+1)}.
\ee
Due to symmetry, recursion relations for $Q_R(k)$ are of the same form as
those for $Q_L(k)$. We therefore now limit ourselves to studying $Q_L(k)$. We
further find it convenient to rescale $Q_L(k)$ as
\bea\label{eq:q_L_def}
q_L(k)&\equiv& \frac{Q_L(k)}{\beta_1\beta_2\cdots\beta_k},
\eea
where $\beta_k = \alpha_k\delta_k$ with $\alpha_k$ and $\beta_k$ defined in
Eqs. (\ref{eq:alpha}) and (\ref{eq:delta}). 
In terms of these rescaled statistical weights, the Garel-Orland recursion
relation\cite{garel_04} becomes:
\be\label{eq:q_L}
\beta_{k+1} q_L(k+1)=q_L(k)+\sigma_{0,k+1}\sum_{j=0}^{k-1}g(k-j)q_L(j)
\ee
for $k=1,...,N$.  We above introduced the local cooperativity parameter
$\sigma_{0,k} = \xi \delta_k$. The ``boundary'' conditions are $q_L(0)=1$ and
$q_L(1)=1/\beta_1$.  We note that, in principle, the recursion relation above
can be used to predict DNA melting curves.  However, the quantity $q_L(k)$
typically ``explodes'' exponentially with the number of basepairs. Therefore, 
two alternative numerical methods, the Poland algorithm and the Fixman-Freire
approximation, are described in appendix \ref{sec:numerics}.

Let us consider the case of homo-DNA, i.e. when the basepair energies are the
same along the DNA. We then set $\beta_k=\beta$ and $\sigma_{0,k}=\sigma_0$ in
Eq. (\ref{eq:q_L}).  We notice that the last term in Eq. (\ref{eq:q_L}) has
the form of a convolution (if we add the term $j=k$ to the sum). We therefore
introduce the generating function (if we replace $z\rightarrow 1/z$ below we
get a $z$-transform)
\be\label{eq:generating_function}
\bar{q}_L(z)=\sum_{k=0}^\infty q_L(k)z^{k}
\ee
which, after using the convolution theorem for $z$-transforms and some
rearrangements (for convenience we also define $g(0)\equiv 0$), gives the
solution to Eq. (\ref{eq:q_L}) as:
\be\label{eq:q_L_s}
\bar{q}_L(z)=\frac{1}{1-\beta^{-1}z[1+\sigma_0\bar{g}(z)]},
\ee
where 
\be\label{eq:g_bar} 
\bar{g} (z)=\sum_{k=1}^\infty g(k)z^k 
\ee 
For the case $g(k) = k^{-c}$ we have that
$\bar{g}(z) = {\rm Li}_c(z)$ is the polylogarithmic function. It is
noteworthy that analytic continuation of the polylogarithmic function to
the complex plane has a branch cut for ${\rm Re} \ z >1$. The
generating function $\bar{q}_L(z)$ therefore also has a branch cut in
the same domain.

\begin{figure}
\begin{center}
\includegraphics[width=8cm]{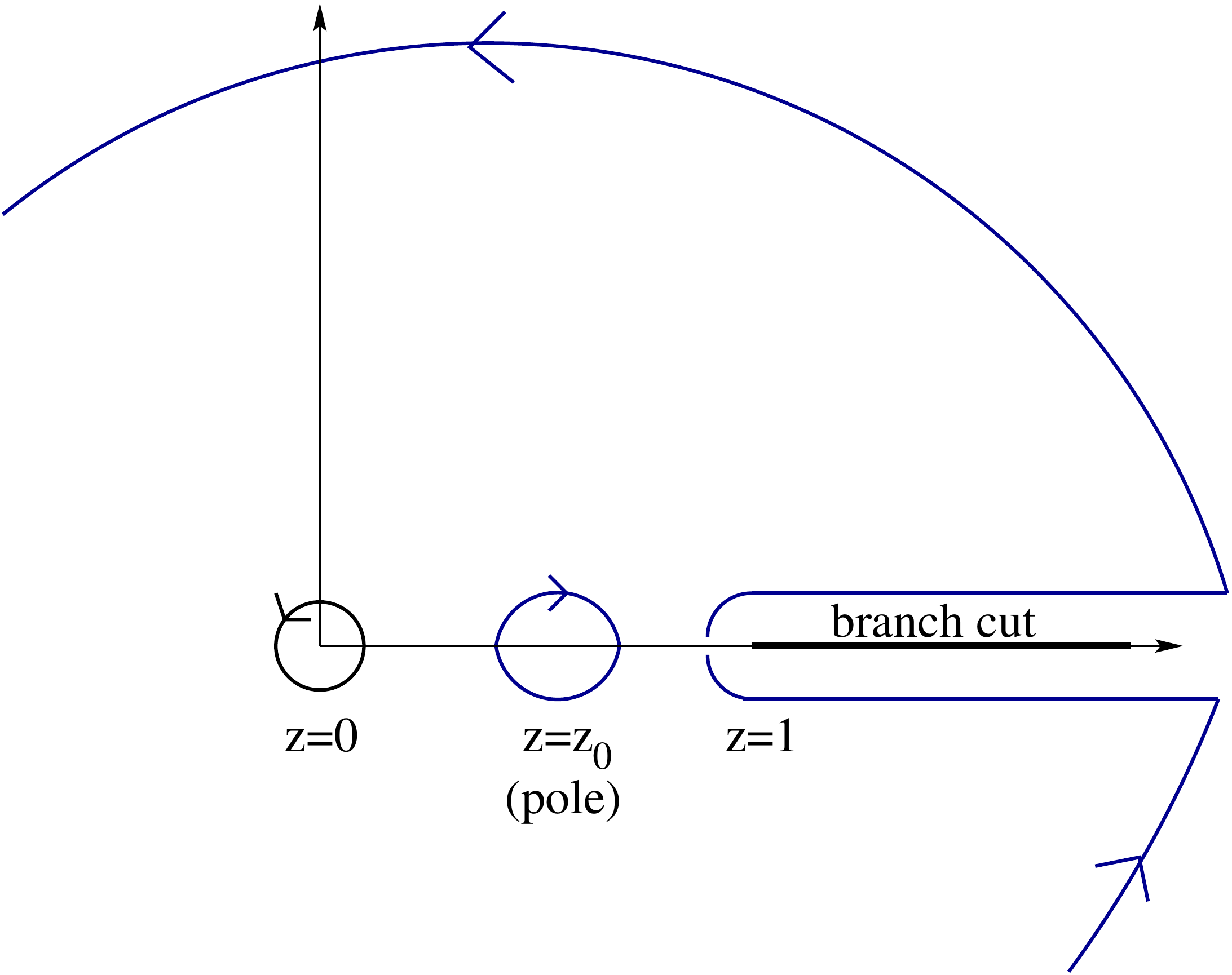}
\end{center}
\caption{Contours used for solving the Poland-Scheraga model
  analytically for homo-DNA. Integration along the black circle around
  $z=0$ gives the statistical weight $q_L(k)$ through
  Eqs. (\ref{eq:q_L_s}) and (\ref{eq:q_L_contour}). The blue contours
  depict a useful deformation of the black contour: We deform the
  contour to ``infinity''.  When
  deforming the original contour, we ``pass'' the pole at $z=z_0$,
  which must be subtracted (notice the ``negative'' orientation of the
  contour around $z=z_0$). The contour along ${\rm Re}\ z = [1,\infty]$
  is needed for the case of a power law form for the
  loop factor, as the associated $z$-transform (polylogarithmic
  function) has a branch cut there. }
\label{fig:contour_int}
\end{figure}

In order to invert the exact result for the generating function,
Eq. (\ref{eq:q_L_s}), we note, see Eq. (\ref{eq:generating_function}), that
$q_L(k)$ is the $k$:th term in a series expansion in $z$, i.e.,
\be\label{eq:q_L_contour}
q_L(k)=\frac{1}{k!}\frac{d^k \bar{q}_L(z)}{d z^k}|_{z=0}=\frac{1}{2\pi 
i}\oint_C 
\frac{\bar{q}_L(z)}{z^{k+1}} dz  
\ee
where the last equality is a standard result in complex analysis.\cite{wunsch} 
The integration above is over a closed contour $C$ around $z=0$, see Fig. 
\ref{fig:contour_int}.
Following Ref. \onlinecite{Wiegel} the contour integral in
Eq. (\ref{eq:q_L_contour}) is evaluated by deforming the original contour as
described in Fig. \ref{fig:contour_int}. The deformed contour has several
parts. One of these parts is a contour around the pole $z_0$ of
$\bar{q}_L(z)$, and, in fact, the thermodynamic behavior is determined by
the contribution from this pole.\cite{Wiegel} For large $k$ we therefore have
\be \label{eq:q_L_deformed} 
q_L(k)\sim \frac{1}{2\pi i}\oint_{C_0}
\frac{\bar{q}_L(z)}{z^{k+1}} dz
\ee 
where $C_0$ is the contour around the pole
depicted in Fig. \ref{fig:contour_int}.  The pole,
$z_0$, is determined by the condition $\bar{q}_L(z_0)=0$, i.e., 
\be \label{eq:z0} 
\bar{g}(z_0)=\frac{\beta-z_0}{\sigma_0 z_0} 
\ee 
where we used Eq. (\ref{eq:q_L_s}). So, using the Cauchy theorem\cite{wunsch} 
we 
have:
\be\label{eq:I1} q_L(k)\sim-{\rm
  Res}[\frac{z^{-k-1}}{1-\beta^{-1}z[1+\sigma_0\bar{g}(z)]},z=z_0]
\ee
where ${\rm Res}[...]$ denotes the residue. Using the fact 
that\cite{wunsch} ${\rm Res}[A(z)/B(z),z=z_0] = A(z_0)/B'(z_0)$ if $z_0$ is a 
simple pole and $A(z)$ is a regular function, we find: 
\be
q_L(k)=\frac{\beta z_0^{-k-1}}{1+\sigma_0[\bar{g}(z_0)+z_0\bar{g}'(z_0)]}.
\ee 
This result holds provided that $z_0\in [0,1]$; if $z_0\ge 1$ then the pole
is in a branch cut (for a power-law form of $g(m)$), see Fig. 
\ref{fig:contour_int}, and does not contribute
to $q_L(k)$.  Finally, we simplify the results above using Eq. (\ref{eq:z0}),
and combining with Eq. (\ref{eq:q_L_def}) we find the partition function:
\be\label{eq:q_L_final}
q_L(k) \sim \frac{z_0^{-k}}{1+\sigma_0z_0\beta^{-1}\bar{h}(z_0)} 
\ee 
where we defined the function: $\bar{h}(z) = z \bar{g}'(z)$.
Eq. (\ref{eq:q_L_final}) is our final expression for the rescaled
partition function for the Poland-Scheraga model for homo-DNA melting
in the thermodynamic limit. For the general case we evaluate $\bar{g}(z)$
using the definition, Eq. (\ref{eq:g_bar}). Using this same definition,
we find that
\be\label{eq:h_bar} 
\bar{h} (z)=\sum_{k=1}^\infty k g(k) z^k 
\ee 
thereby providing an explicit expression for all quantities in
Eq. (\ref{eq:q_L_final}).

Finally, by combining Eqs. (\ref{eq:P_open}), (\ref{eq:q_L_def}), and
(\ref{eq:q_L_final}), we obtain our final expression for the melting 
probability:
\be\label{eq:P_zl1} P =1-\frac{1}{1+\sigma_0 \beta^{-1}z_0 \bar{h}(z_0)} 
\ee 
where $\bar{g}(z)$ is given in Eq. (\ref{eq:g_bar}) and $\bar{h}(z)$
in Eq. (\ref{eq:h_bar}). These expressions can be used to numerically
evaluate $\bar{g}(z)$ and $\bar{h}(z)$ by replacing the series by sums. The
quantity $z_0$ is determined by (numerically) solving Eq. (\ref{eq:z0}). Thus, 
the equations given here provide means for straightforward calculation of the 
melting probability for different choices
of $g(m)$ in the thermodynamic limit. For the case $g(k) = k^{-c}$ we have
$\bar{g}(z) = {\rm Li}_c(z)$ and $\bar{h}(z) = {\rm Li}_{c-1}(z)$, where $
{\rm Li}_c(z)$ is the polylogarithmic function.

\section{Analysis of melting curve for unconfined DNA 
}\label{sec:unconfined_DNA}

This appendix provides results for melting curves of unconfined DNA,
and, thus, serves a background to the main text, where melting of
confined DNA molecules is studied.

\subsection{Deviation from the Marmur-Doty melting temperature}
\label{sec:deviation_melting_temp}

The Marmur-Doty formula, Eq. (\ref{eq:T_MD}), is based on a simple
estimate where one assumes melting to occur at a temperature where the
free energy for breaking one hydrogen bond and one stacking
interaction equals zero. Due to, for instance, neglect of the
``boundary energy'' $(1/2)R T \log \xi$ and the loop factor
$g(m)$, which favors smaller bubbles, Eq. (\ref{eq:T_MD}) is only an
approximation.

\begin{figure}
\label{fig:NB_random_c_theo}
  \begin{center}
    \includegraphics[width=8.5cm]{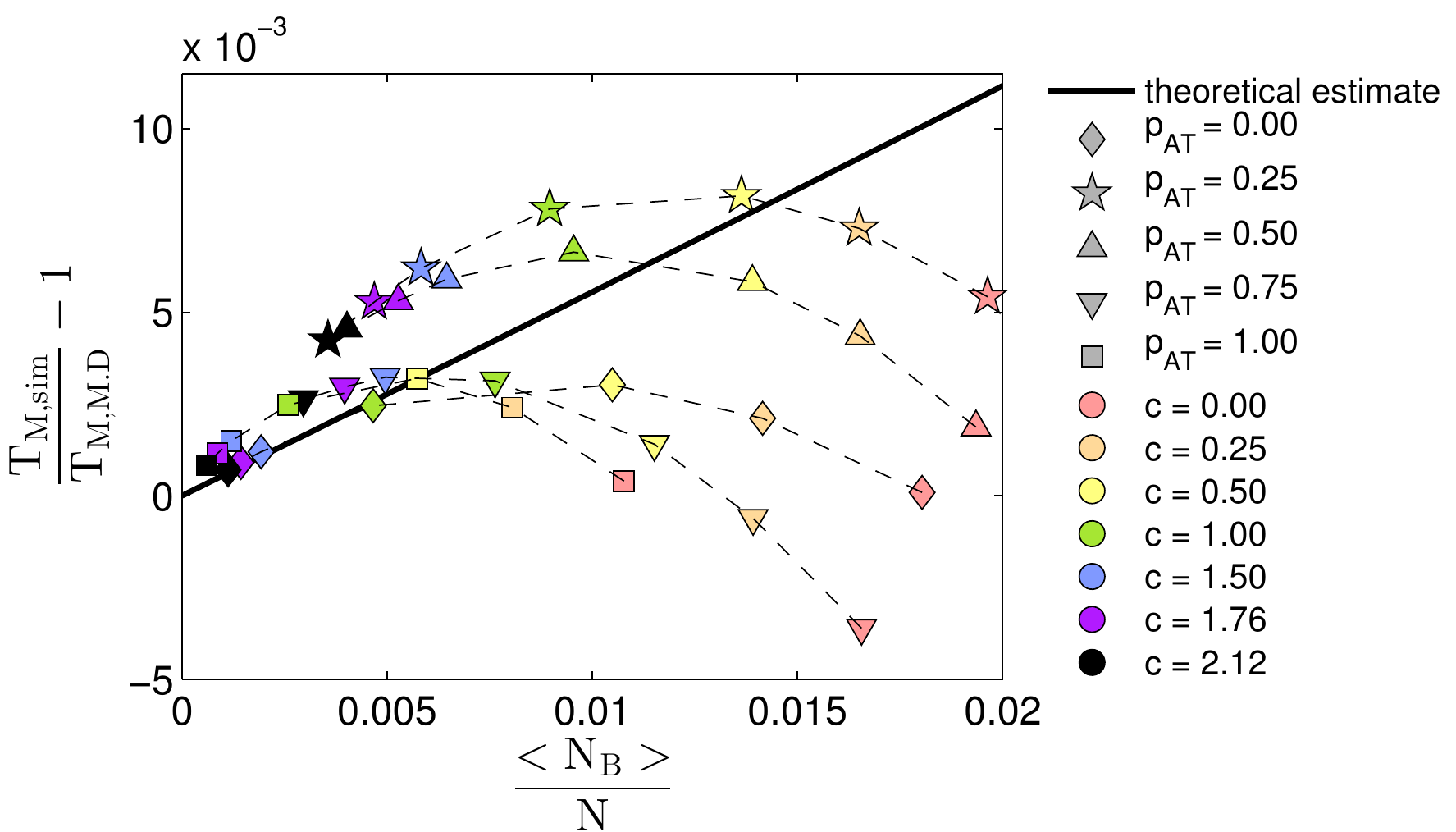}
  \end{center}
  \caption{Shift of the simulated melting temperature $T_{\rm
      {M,sim}}$  (units of  Kelvin) compared to the Marmur-Doty melting 
temperature
    $T_{\rm {M,M.D.}}$ as a function of the simulated mean number of
    boundaries at $T_{\rm {M,sim}}$ at $T_{\rm {M,sim}}$ for random unconfined 
    DNA. The solid line shows our theoretical estimate from
    Eq. (\ref{eq:TM_NB}).  For $c\geqslant 1$, there is clear relation
    between the number of bubbles and the deviation from the
    Marmur-Doty melting temperature. For larger values of $c$,
    however, we find less correlation between the shift in melting
    temperature and $\langle N_B \rangle $. Averages are taken over
    200 different sequences of length $2\cdot 10^5$ basepairs at AT
    ratios $p_{\rm{AT}}={0,0.25,0.5,0.75,1}$ (different markers), each
    for different loop exponents $c={2.12,1,76,1.5,0.5,0}$ (indicated
    by different colors).}
\label{fig:NB_random_c}
\end{figure}

In order to estimate shifts in melting temperature compared to the
Marmur-Doty formula for unconfined DNA, we note that an entropic
contribution $R 
\log \xi$
has to be assigned to each bubble. The corrected melting
temperature can then be estimated from the number of bubbles, $N_B$, as
\be\label{eq:TM_NB} 
T_{\rm{M}}^{\rm B}=\frac{\sum H}{\sum
  S}=\frac{S_{\rm{ref}}}{S_{\rm{ref}}-R \log \xi \frac{N_{\rm B}}{N}
}T_{\rm M} 
\ee 
In practice, we further replace $N_B$ by its expected number of
bubbles $\langle N_B \rangle$. In Figure \ref{fig:NB_random_c} we
tested the above mean-field prediction in the following way: using the
Fixman-Freire approximation we determined the shift in melting
temperature from the Marmur-Doty formula for a range of $p_{\rm AT}$
and loop exponents $c$. Simultaneously, we determined the expected
number of bubbles, $\langle N_B \rangle$, see Section
\ref{sec:numerics} for details. From Figure \ref{fig:NB_random_c} we
notice that, while Eq. (\ref{eq:TM_NB}) captures the typical behavior
for the shift as a function of number of bubbles, there are
deviations.  For $c\geqslant 1$, there is clear relation between the
number of bubbles and the deviation from the Marmur-Doty melting
temperature. For larger values of $c$, however, the trend is less
pronounced. A possible explanation for this ``large $c$'' effect is that while
the curves get wider with smaller values of $c$, the asymmetry of the
melting transition results in the point $T(f=0.5)$ to be shifted to
smaller temperatures.

Notice that we in the main text avoid ``problems'' with the
Marmur-Doty formula by always using simulated melting temperatures as
reference.

A final word of caution is here in order. The hydrogen bond energies
we use in the simulations are not measured for individual base-pairs,
but estimated from global melting of a few model sequences and
therefore already contain the contribution of the ring factor. We
therefore get too large temperatures from our simulations already for
unconfined DNA with $c=2.12$, where the Marmur-Doty relation was found 
experimentally.\cite{owen,FK} For smaller values of c (or smaller 
channels), the
melting transition gets broader and the mean number of bubbles at
$T_M$ increases.

\subsection{Width of the melting transition as a function of loop exponent c}

\begin{figure}
  \begin{center}
    \includegraphics[width=8cm]{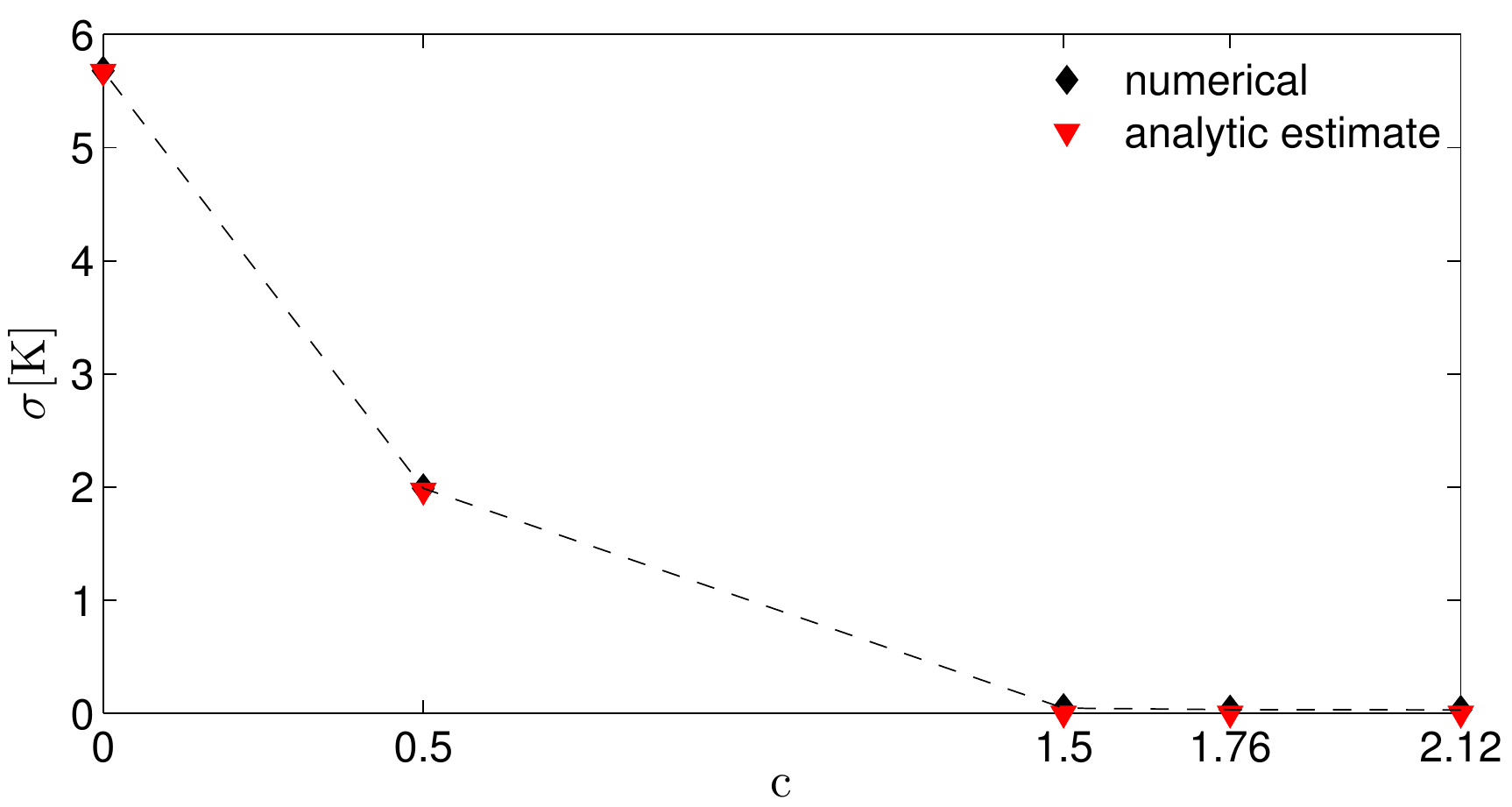}
    \includegraphics[width=8cm]{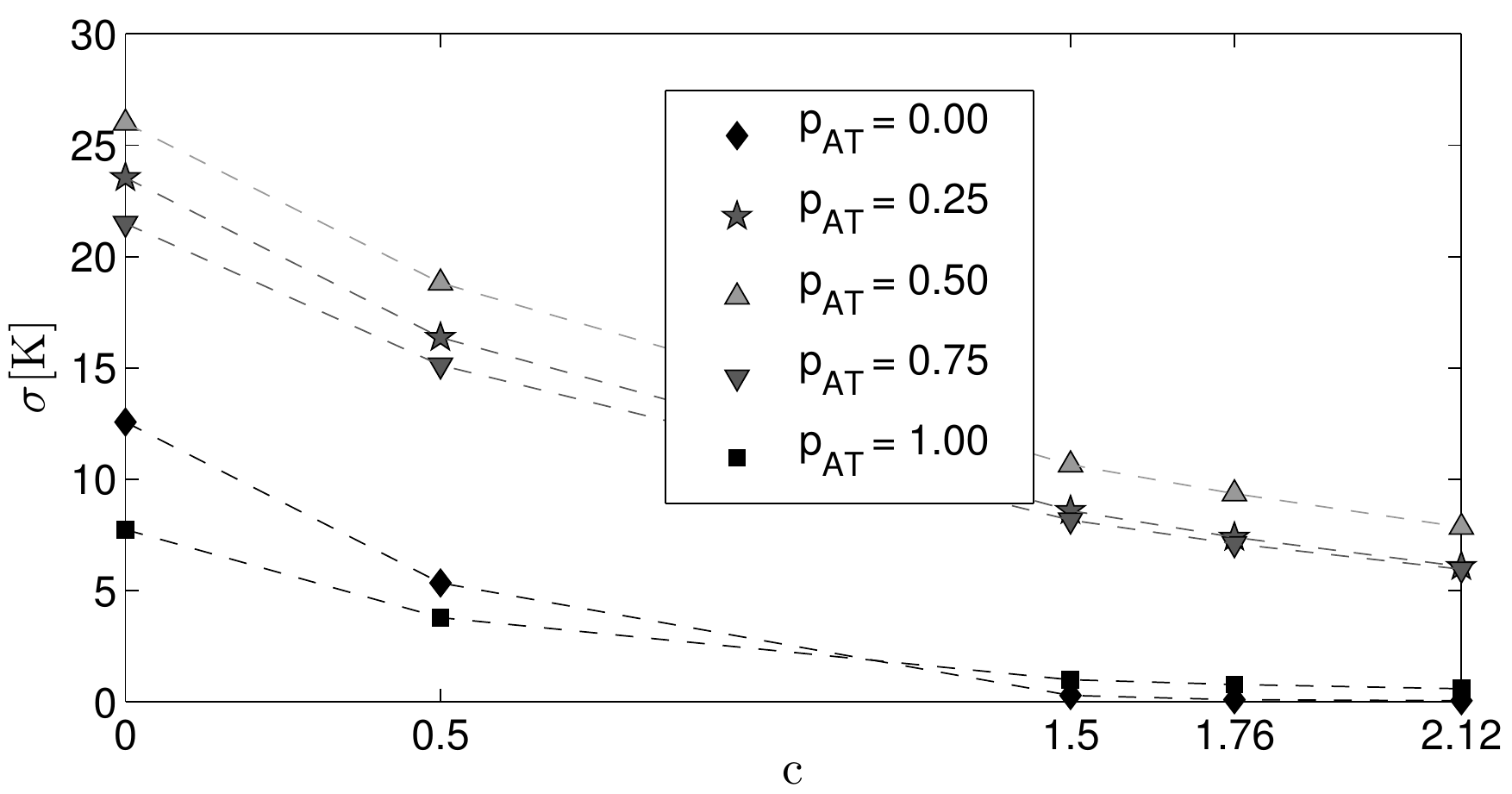}
     \includegraphics[width=8cm]{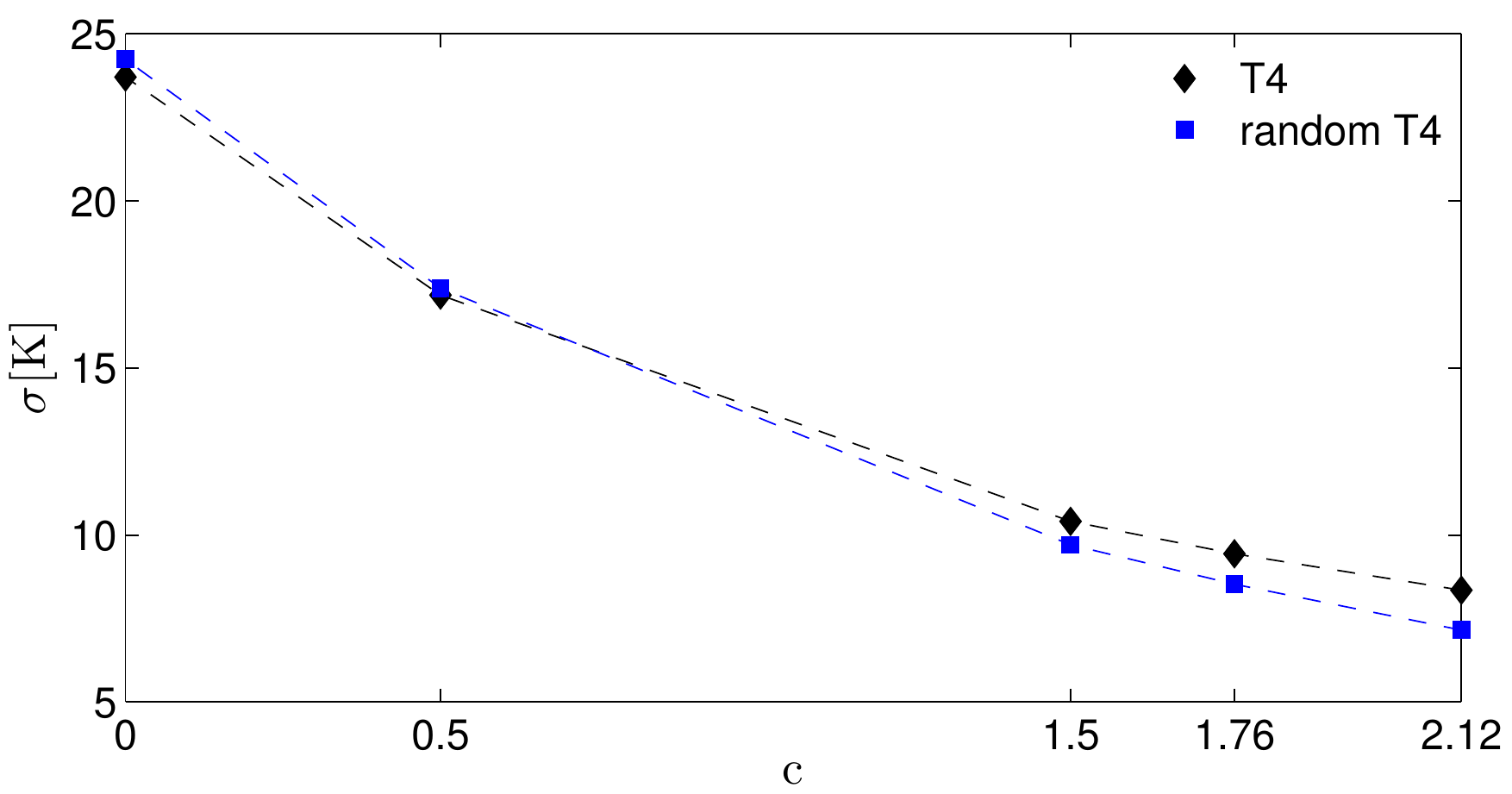}
  \end{center}
\caption{Width of the melting curve as a function of the loop exponent $c$ for 
unconfined homo-DNA (top), unconfined random DNA (middle) and
 unconfined T4-phage DNA (bottom) (with associated 
random DNA, same AT fraction). Parameters are the same as in Fig. 
\ref{fig:melt_random}.}
\label{fig:width_vs_c}
\end{figure}

Figure \ref{fig:width_vs_c} shows the change of the width $\sigma$ for
different loop exponents c for unconfined DNA. The width was extracted
using the same approach as described in subsection
\ref{sec:melt_homo}. We find that increasing $c$ decreases the width of
the transition as it should. These findings hold for the three cases
of interest here: 1. homo-DNA, 2. random DNA for different AT
fractions, and 3. T4 phage DNA. The prefactors for the width are larger
for case 2. than for case 1. The homo-DNA case does not give identical results 
to the $f_{\rm AT}=0$ and $f_{\rm AT}=1$ cases due to the additional different
stacking parameters for the latter cases as compared to the homo-DNA case, see 
section \ref{sec:Marmur-Doty}.

\subsection{Analytical prediction for different loop exponents c}

\begin{figure}
  \begin{center}
    \includegraphics[width=8cm]{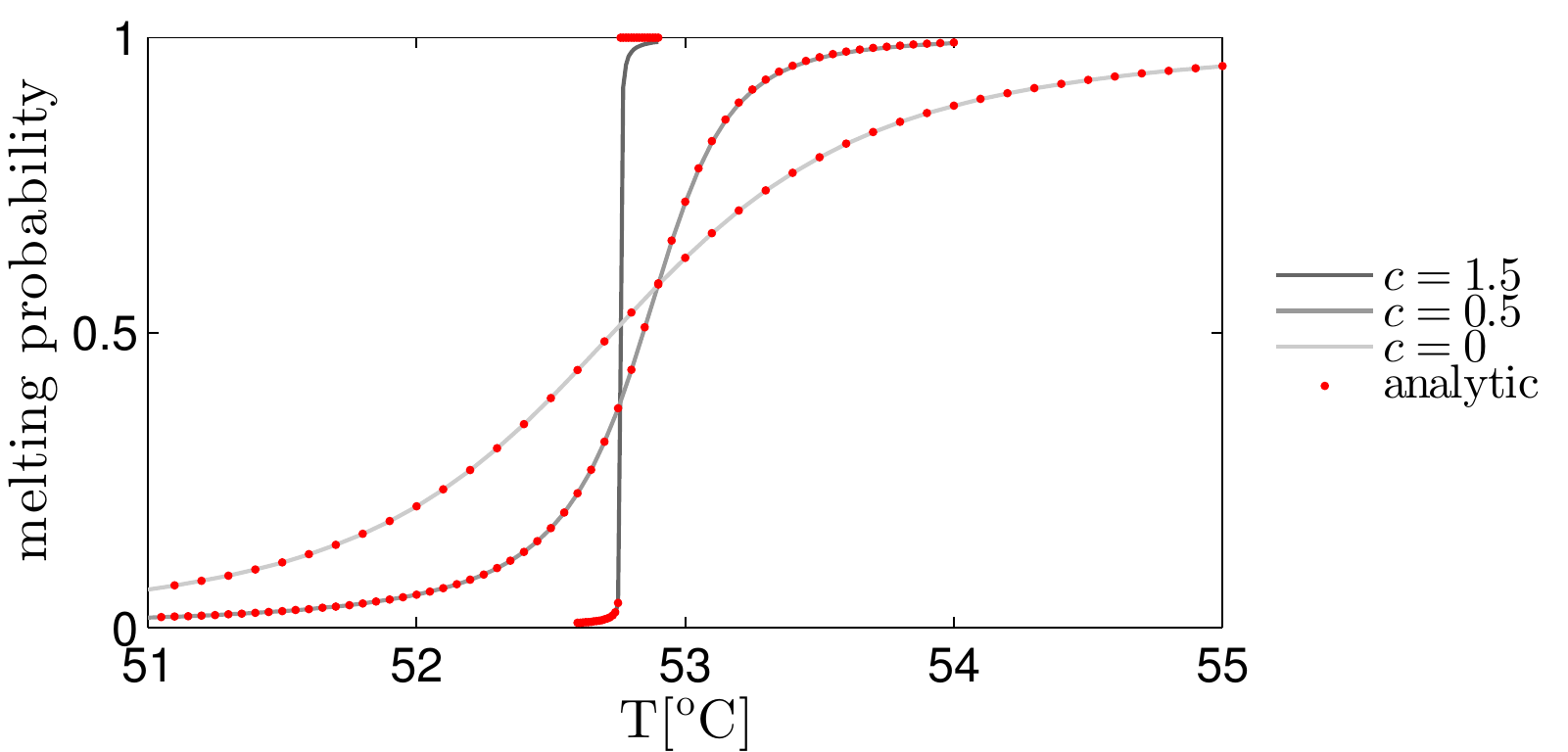}
  \end{center}
  \caption{Melting probability as a function of temperature for an unconfined
    homo DNA for different loop exponents. The numerical results for a
    sequence of length 50 kilo basepairs (solid lines) are compared to the
    analytic prediction in the thermodynamic limit (red marks). We used $\xi =
    10^{−3}$ and melting temperature for a sequence `A ... A', see
    Eq. (\ref{eq:TM_A}).}
\label{fig:melting_homo_c}
\end{figure}

Figure \ref{fig:melting_homo_c} shows the melting probability for unconfined
homo-DNA for different values of the loop exponent $c$, both using the
numerical Poland algorithm and the analytic prediction introduced in section
\ref{sec:homo_DNA}.  As before, we used Eq. (\ref{eq:P_zl1}), where $z_0$ was
numerically obtained by solving Eq. (\ref{eq:z0}). We find very good
agreement between the numerical and analytical result. Note, however, that due
to the finite size of the DNA sequences the numerical result for $c=1.5$
deviates slightly from the analytical result, which is obtained for the
thermodynamic limit.

\end{document}